# Momentum-scalar coupled turbulence with anomalous momentum and scalar diffusions. Part 2: With short-range external force and implementation in electrokinetic turbulence

Wei Zhao (赵伟)

*State Key Laboratory of Photon-Technology in Western China Energy, International Scientific and Technological Cooperation Base of Photoelectric Technology and Functional Materials and Application, Laboratory of Optoelectronic Technology of Shaanxi Province, Institute of Photonics and Photon-technology, Northwest University, Xi'an 710127, China*

**Abstract** We extend the generalized anomalous diffusion framework established in Part 1 to the short-range forcing regime ($\beta > 2/3$), where the multiscale-force dominated (MFD) subrange is intercalated after the inertial subrange, competing directly with the dissipation ranges. Focusing on electrokinetic (EK) turbulence as the prototypical example with $\beta = 1$, we derive the relations for the velocity and scalar dissipation wavenumbers, $k_{MD}$ and $k_{SD}$, across all four subranges of the Quad-cascade process (inertial, constant-$\Pi_u$, constant-$\Pi_s$, and variable flux). By incorporating Golestanian's predicted anomalous diffusion regimes for electrolytes, we construct comprehensive phase diagrams showing how the relative magnitudes of $k_{MD}$ and $k_{SD}$ are governed by the scale-dependent anomalous Schmidt number $Sc_Z$. We identify two new spectral subranges that emerge exclusively in this short-range forcing regime: (i) the convective-forcing-viscous subrange ($k_{MU} \ll k \ll k_{CFVD}$) for either $Sc_Z \gg 1$ or $Sc_Z \ll 1$ with $\gamma > \alpha$, where the scalar field drives viscous flow through electric body force; and (ii) the diffusive-forcing-convective subrange ($k_{MU} \ll k \ll k_{DFCD}$) for $Sc_Z \ll 1$ and $\gamma \le \alpha$, where the scalar structure dominated by diffusion determines electric body force, which in turn cause momentum convection. These results provide a complete analytical map of EK turbulence under anomalous diffusion, revealing how external parameters such as electric field strength and ionic diffusivity determine the cascade topology.

## 1. Introduction

In the classical picture of fully developed turbulence, turbulent kinetic energy injected at large scales undergoes a forward cascade through the inertial subrange before being dissipated by ordinary molecular diffusion at small scales. This paradigm is applicable to a broad area of physical phenomena, including free-decaying turbulence [1,2], buoyancy-driven turbulence [3,4] and even some quantum turbulence regimes [5,6] where the classical cascade picture provides an approximate description at sufficiently large scales. However, a wide class of natural and engineering turbulent flows—particularly in electrokinetic phenomena [7,8], polymer science [9], and active matter [10,11]—are driven by external forces that act predominantly at intermediate or small scales. In these phenomena, the force amplitude is concentrated at intermediate to high wavenumbers, fundamentally altering the cascade topology.

As established in Part 1 [12] of this two-part series, the fractional order $\beta$ of the external force $\boldsymbol{M}\mathfrak{D}_{(1)}^{\beta/4}\widehat{s'}$ provides a rigorous criterion for distinguishing these two scenarios. When $\beta < 2/3$, the force acts at large scales (long-range forcing), and the forward cascade picture is retained. When $\beta > 2/3$, however, the force becomes important at small scales, corresponding to short-range forcing [13,14]. In this case, the classical picture of a unidirectional forward cascade breaks down entirely: the multiscale-force dominated (MFD) subrange shifts to the high-wavenumber side of the inertial subrange, and energy injected at this subrange propagates both towards low wavenumbers (inverse cascade) and high wavenumbers (forward cascade). This topological bifurcation, termed the Quad-cascade process [14-16], features four distinct sub-branches—inertial, constant-$\Pi_u$, constant-$\Pi_s$, and variable flux—each with different scaling exponents in turbulent kinetic energy and scalar spectra.

In Part 1, we established the generalized spectral framework for momentum–scalar turbulence with arbitrary anomalous diffusions of momentum and scalar for no external forcing or long-range forcing ($\beta < 2/3$). In those cases, anomalous diffusions merely modify dissipations (characteristic wavenumber and spectra) within an otherwise classical forward-cascade topology. The present paper (Part 2) addresses a fundamentally different question, i.e. how do anomalous diffusions reshape the already complex Quad-cascade topology when the forcing is short-range ($\beta > 2/3$)? Specifically, does the anomalous Schmidt number $Sc_Z = k_0^{\gamma-\alpha} c_u / c_s$ govern whether the scalar dissipation wavenumber $k_{SD}$ falls below or above the kinetic dissipation wavenumber $k_{MD}$? And what new spectral subranges emerge from the competition between fractional forcing, fractional diffusion, and nonlinear convection in this high-$k$ forcing regime?

We further focus on EK turbulence [14-16] as the paradigmatic example of short-range forcing. In EK flow, the electric body force (EBF) generated under an external electric field is highly dependent on the control scalars—electric permittivity ($\epsilon$) and electric conductivity ($\sigma$)—through the local electric field $E = E(x, t; \sigma, \epsilon)$

$$\boldsymbol{F}_e = (\nabla \cdot \varepsilon \boldsymbol{E})\boldsymbol{E} - \frac{1}{2}(\boldsymbol{E} \cdot \boldsymbol{E})\nabla\varepsilon + \frac{1}{2}\nabla\left[\left(\frac{\partial \varepsilon}{\partial \rho}\right)_T \rho \boldsymbol{E} \cdot \boldsymbol{E}\right] \tag{1}$$

Under a series of hypotheses on the velocity and scalar fields, including homogeneous, isotropic, small perturbations and etc, the EBF applied on the turbulent flow corresponds to $\beta = 1$ — a canonical case that lies well within the short-range regime. The scalar can be $\sigma$, $\epsilon$ or other scalars that linearly related to these two physical quantities. For ordinary diffusion, the momentum transport is described by the Navier–Stokes equation coupled with scalar transport equations for $\sigma$ and $\varepsilon$. However, as recently shown by Golestanian [17], electrolytes under strong external electric fields exhibit anomalous diffusion, transitioning from a ballistic regime ($\alpha = 1$, fractional order of anomalous scalar diffusion [12]) through a medium-time anomalous regime ($\alpha = 4/3$) to a long-time anomalous regime ($\alpha = 3/2$) depending on the time and mean-square displacement of observation. The interplay among these multiple $\alpha$ values and the four Quad-cascade sub-branches generate an exceptionally rich phenomenology that cannot be captured by classical theories.

The main contributions of this Part 2 are threefold. First, we extend the turbulent flow with external forcing to $\beta > 2/3$, deriving the relations for $k_{MD}$ and $k_{SD}$ for all four Quad-cascade sub-branches. Second, we specialize to EK turbulence ($\beta = 1$) and incorporate the three distinct anomalous diffusion phases of ions, constructing comprehensive phase diagrams that map the cascade topology as a function of external parameters. Third, we identify two novel spectral subranges — the convective-forcing-viscous (CFV) subrange and the diffusive-forcing-convective (DFC) subrange — that exist exclusively in the short-range forcing regime and have no counterpart in Part 1 or in classical theories.

The paper is organized as follows. In section 2, we succinctly restate the key fractional conservation laws from Part 1, keeping only the governing equations for self-consistency. In section 3, we present the general solutions for $k_{MD}$ and $k_{SD}$ in the short-range forcing regime. In section 4, we specialize to EK turbulence ($\beta = 1$), incorporating the three anomalous diffusion phases of electrolytes and constructing the full phase diagram. In section 5, we derive the characteristic wavenumbers of EK turbulence regarding anomalous diffusions of momentum and scalar. Then, the corresponding velocity and scalar spectra in CFV and DFC subranges are theoretically analyzed in section 6 and summarized in section 7 finally.

## 2. Conservative laws and models

In this investigation, the governing equations of velocity ($\widehat{\boldsymbol{u}}$) and scalar fluctuations ($\widehat{s'}$) regarding anomalous diffusions in the turbulence driven by the multiscale force can be expressed as [12]

$$\left(\frac{\partial}{\partial t} + \widehat{\boldsymbol{u}} \cdot \boldsymbol{\nabla}\right) \widehat{\boldsymbol{u}} = -\frac{1}{\rho} \nabla \hat{p} + c_u \mathfrak{D}_{(2)}^{\frac{\gamma}{4}} \widehat{\boldsymbol{u}} + \boldsymbol{M} \mathfrak{D}_{(1)}^{\frac{\beta}{4}} \widehat{s'} \qquad (2a)$$

$$\left(\frac{\partial}{\partial t} + \widehat{\boldsymbol{u}} \cdot \boldsymbol{\nabla}\right) \widehat{s'} = -\boldsymbol{N} \cdot \widehat{\boldsymbol{u}} + c_s \mathfrak{D}_{(2)}^{\frac{\alpha}{4}} \widehat{s'} \qquad (2b)$$

$$\boldsymbol{\nabla} \cdot \widehat{\boldsymbol{u}} = \boldsymbol{0} \qquad (2c)$$

where $\rho$ is the referenced fluid density, $\hat{p} = \hat{p}(\boldsymbol{x}, t)$ is pressure, $\boldsymbol{M} \mathfrak{D}_{(1)}^{\beta/4} \widehat{s'}$ is the multiscale force related to the scalar field $\widehat{s'}$ [13,14], with $\boldsymbol{M}$ being the dimensional vector associated with the physical field. $\boldsymbol{N} = \boldsymbol{\nabla} \langle \hat{s} \rangle$ is the vector related to the mean scalar $\langle \hat{s} \rangle$ to characterize the feature of stratified scalar background. $\mathfrak{D}_{(1)}^{\chi/4}$ and $\mathfrak{D}_{(2)}^{\chi/4}$ are the first and second subsets of the fractional biharmonic operator of order $\chi/4$ (see Part 1 [12]). For the scalar field, it is assumed that $\mathfrak{D}_{(2)}^{\alpha/4} \langle \hat{s} \rangle \ll \mathfrak{D}_{(2)}^{\alpha/4} \widehat{s'}$. The diffusion coefficients $c_u$ and $c_s$ have dimensions $L^{\gamma} T^{-1}$ and $L^{\alpha} T^{-1}$, respectively, and reduce to kinematic viscosity and scalar diffusivity when $\gamma = \alpha = 2$.

In Fourier space, the governing equations of turbulent kinetic energy and scalar variance are [14,18-21]

$$\frac{\mathrm{d}}{\mathrm{dt}} \tilde{E}_u(\boldsymbol{k}) = \tilde{T}_u(\boldsymbol{k}) + \tilde{D}_u(\boldsymbol{k}) + \tilde{F}_s(\boldsymbol{k}) \qquad (3a)$$

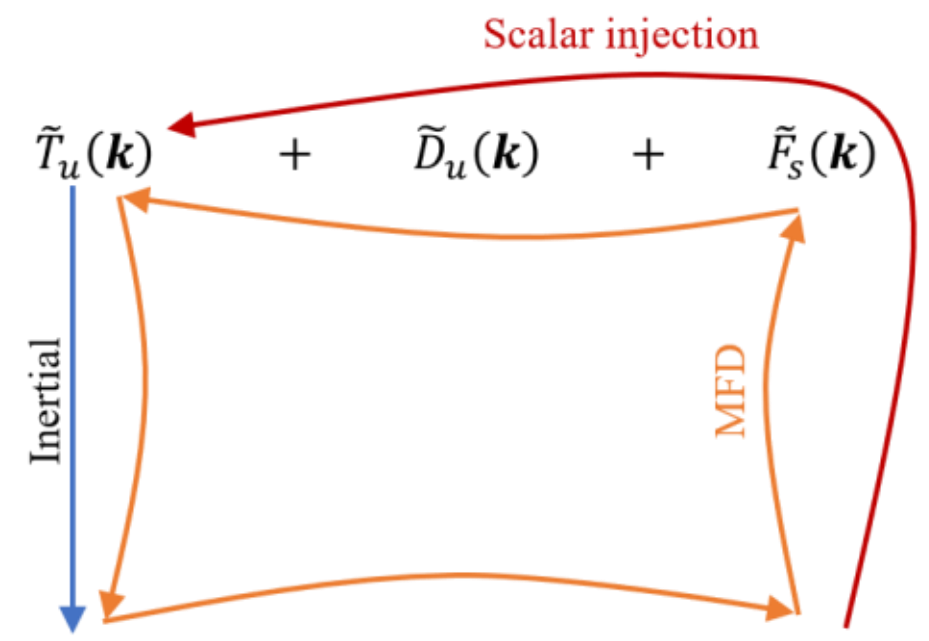

$$\frac{\mathrm{d}}{\mathrm{dt}} \tilde{E}_s(\boldsymbol{k}) = \tilde{T}_s(\boldsymbol{k}) + \tilde{D}_s(\boldsymbol{k}) - \tilde{F}_A(\boldsymbol{k}) \qquad (3b)$$

where $\tilde{E}_u(\boldsymbol{k}) = |\boldsymbol{u}(\boldsymbol{k})|^2/2$ and $\tilde{E}_s(\boldsymbol{k}) = |s'(\boldsymbol{k})|^2/2$ are the modal kinetic energy and scalar variance. $\tilde{T}_u(\boldsymbol{k})$ and $\tilde{D}_u(\boldsymbol{k})$ are the nonlinear kinetic energy transfer rate and dissipation rate corresponding to anomalous momentum diffusion, respectively. $\tilde{T}_s(\boldsymbol{k})$ and $\tilde{D}_s(\boldsymbol{k})$ are the nonlinear transfer rate of the scalar variance and scalar dissipation rate corresponding to anomalous scalar diffusion, respectively. $\tilde{F}_s(\boldsymbol{k})$ denotes the energy feeding rate by the multiscale force due to the scalar field and $\tilde{F}_A(\boldsymbol{k})$ is the scalar feeding rate by bulk components. In Fourier space, the spectral budget equations under statistical equilibrium read

$$\frac{\mathrm{d}}{\mathrm{d}k}\Pi_u(k) = F_s(k) + D_u(k) \tag{4a}$$

$$\frac{\mathrm{d}}{\mathrm{d}k}\Pi_s(k) = -F_A(k) + D_s(k) \tag{4b}$$

where $F_s(k)$ and $F_A(k)$ are the 1D energy feeding rate by multiscale force due to the scalar field and the scalar feeding rate at wavenumber $k$, $D_u$ and $D_s$ are the corresponding dissipation terms respectively. When $dk \to 0$, they are

$$F_s(k) = k^{\beta} \sum_{k<|\boldsymbol{k}'|\le k+dk} \mathrm{Re}[s'(\boldsymbol{k}')\boldsymbol{M} \cdot \boldsymbol{u}^*(\boldsymbol{k}')] \tag{5a}$$

$$F_A(k) = \sum_{k<|\boldsymbol{k}'|\le k+dk} \mathrm{Re}[s'(\boldsymbol{k}')\boldsymbol{N} \cdot \boldsymbol{u}^*(\boldsymbol{k}')] \tag{5b}$$

$$D_u(k) = 2c_u k^{\gamma} \mathrm{Re}\left(e^{i\frac{1}{2}\pi\gamma}\right) E_u(k) \tag{5c}$$

$$D_s(k) = 2c_s k^{\alpha} \mathrm{Re}\left(e^{i\frac{1}{2}\pi\alpha}\right) E_s(k) \tag{5d}$$

The influence of $\beta$ on the structures of turbulence scaling subranges can be found in Fig. 1. When $\beta > 2/3$, the MFD subrange locates on the high-wavenumber side of the inertial subrange, i.e. $k \gg k_{MI}$, with $k_{MI}$ being the intersection point of inertial and MFD subranges. The turbulent kinetic energy is injected in the $k$-space at a middle-to-high wavenumber regime. It spreads toward low and high $k$ regimes simultaneously, to form inertial subrange through inverse cascade and to consume turbulent kinetic energy in the dissipation subrange. According to Quad-cascade process model [14,16], in the MFD subrange, the scaling properties can be highly diverse, relying on $\beta$. The interaction between MFD subrange and dissipation subrange is controlled by both $\beta$ and $\gamma$, showing dependence on characteristic wavenumber $k_{MD}$ (will be discussed in Section 3 and 5). Similar issues can also be predicted for the cascade of scalar variance which is dependent on characteristic wavenumber $k_{SD}$. Specifically, the characteristic length scales and their relationship with $\gamma$ and $\alpha$ are derived.

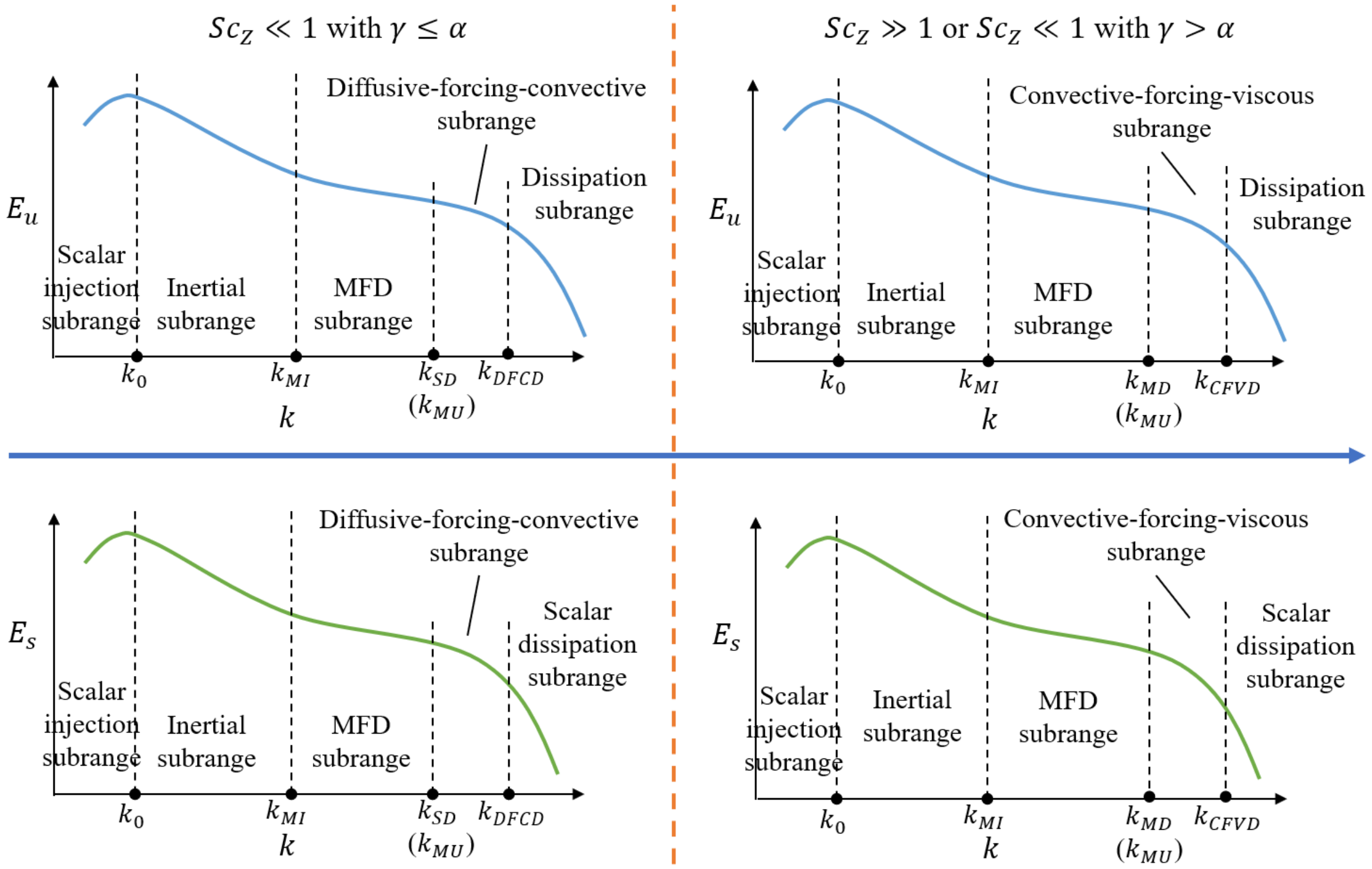

Fig. 1 Fine structures of momentum-scalar coupling turbulence when $\beta > 2/3$. The definitions of $k_0$, $k_{MI}$, $k_{MD}$, $k_{SD}$, $k_{MU}$ (equals to the smaller one of $k_{MD}$ and $k_{SD}$), $k_{DFCD}$ and $k_{CFVD}$ are defined in the following sections.

## 3. Characteristic scales

In the turbulent driven by the multiscale force $\boldsymbol{M}\mathfrak{D}_{(1)}^{\beta/4}\widehat{s'}$, the overall turbulent kinetic energy is primarily contributed by the injection from external forcing, while finally dissipated. As the major subrange for energy injection, the MFD subrange lies between the inertial subrange and the dissipation subrange at $\beta > 2/3$. The cascade of turbulent kinetic energy is associated with turbulent energy dissipation rate ($\varepsilon_u$). On the other hand, the multiscale force is strictly related to the scalar structures, which is associated with the scalar cascade. Therefore, both $\varepsilon_u$ and scalar dissipation rate ($\varepsilon_s$) dominates the cascade of turbulent kinetic energy. This is a major difference between the inertial subrange and the MFD subrange. Regarding the coupling between momentum and scalar transport, the scalar cascade is also dominated by $\varepsilon_u$ and $\varepsilon_s$. In the MFD subrange, the influence of anomalous diffusions of both momentum and scalar are absent. Accordingly, Eqs. (4) can be rewritten as [14]

$$\frac{\mathrm{d}}{\mathrm{d}k}\Pi_u(k) = F_s(k) \tag{6a}$$

$$\frac{\mathrm{d}}{\mathrm{d}k}\Pi_s(k) = -F_A(k) \tag{6b}$$

For $\boldsymbol{M} \parallel \boldsymbol{N}$, Eqs. (5a, b) give $F_s(k) - \frac{M}{N}F_A(k)k^{\beta} = 0$. Thus, we have the conservation law

$$\frac{\mathrm{d}}{\mathrm{d}k}\Pi_u(k) + \frac{M}{N}k^{\beta}\frac{\mathrm{d}}{\mathrm{d}k}\Pi_s(k) = 0 \tag{7}$$

where $\Pi_u(k) = ku_k^3$ and $\Pi_s(k) = ks_k^2u_k$ [21], $u_k$ and $s_k$ represent the 1D spectral components of velocity and scalar. Besides, another phenomenology relationship can be deduced by balancing the convection term and the $\boldsymbol{M}\mathfrak{D}_{(1)}^{\beta/4}\widehat{s'}$ forcing term in Eq. (2a) as

$$ku_k^2 = k^{\beta}Ms_k \tag{8}$$

Superficially, we can assume $E_u$ and $E_s$ in the MFD subrange are only determined by $\varepsilon_u$, $\varepsilon_s$ and $k$, as $E_u = \varepsilon_u^{a1}\varepsilon_s^{a2}k^{\xi_u}$ and $E_s = \varepsilon_u^{b1}\varepsilon_s^{b2}k^{\xi_s}$. These give the scaling coefficients $a1 = (3+\xi_u)/2$, $a2 = -(5+3\xi_u)/2$, $b1 = (\xi_s+1)/2$ and $b2 = -3(\xi_s+1)/2$. Regarding in the MFD subrange, $E_u \sim u_k^2/k$ and $E_s \sim s_k^2/k$. The expressions of $u_k$ and $s_k$ with given $\xi_u$ and $\xi_s$ can be

$$u_k = \varepsilon_u^{\frac{3+\xi_u}{4}}\varepsilon_s^{-\frac{(5+3\xi_u)}{4}}k^{\frac{\xi_u+1}{2}} \tag{9a}$$

$$s_k = \varepsilon_u^{\frac{\xi_s+1}{4}}\varepsilon_s^{-\frac{3(\xi_s+1)}{4}}k^{\frac{\xi_s+1}{2}} \tag{9b}$$

The scaling properties of the spectral components of velocity and scalar, and their strengths are dominated by $\xi_u$, $\xi_s$, $\varepsilon_u$ and $\varepsilon_s$. Besides, since the MFD subrange lies between the inertial subrange and the dissipation subrange at $\beta > 2/3$, the boundaries of MFD subrange can be estimated as following.

(1) The lower bound of the MFD subrange ($k_{MI}$)

According to the Eqs. (9a, b), in the MFD subrange we have

$$E_u = C_{ZK}\varepsilon_u^{\frac{3+\xi_u}{2}}\varepsilon_s^{-\frac{(5+3\xi_u)}{2}}k^{\xi_u} \qquad (10a)$$

$$E_s = C_{ZP}\varepsilon_u^{\frac{\xi_s+1}{2}}\varepsilon_s^{-\frac{3(\xi_s+1)}{2}}k^{\xi_s} \qquad (10b)$$

where $C_{ZK}$ and $C_{ZP}$ are coefficients to be determined. While in the inertial subrange, $E_u = C_K\varepsilon_u^{\frac{2}{3}}k^{-\frac{5}{3}}$ and $E_s = C_p\varepsilon_u^{-\frac{1}{3}}\varepsilon_s k^{-\frac{5}{3}}$. The characteristic wavenumber $k_{MI}$ where the inertial subrange and MFD subrange intersect can be determined from two perspectives, which are

$$C_K\varepsilon_u^{\frac{2}{3}}k_{MI}^{-\frac{5}{3}} = C_{ZK}\varepsilon_u^{\frac{3+\xi_u}{2}}\varepsilon_s^{-\frac{(5+3\xi_u)}{2}}k_{MI}^{\xi_u} \rightarrow k_{MI} = C_K^{\frac{1}{\xi_u+\frac{5}{3}}}C_{ZK}^{-\frac{1}{\xi_u+\frac{5}{3}}}\varepsilon_u^{-\frac{1}{2}}\varepsilon_s^{\frac{3}{2}} \qquad (11a)$$

$$C_p\varepsilon_u^{-\frac{1}{3}}\varepsilon_s k_{MI}^{-\frac{5}{3}} = C_{ZP}\varepsilon_u^{\frac{\xi_s+1}{2}}\varepsilon_s^{-\frac{3(\xi_s+1)}{2}}k_{MI}^{\xi_s} \rightarrow k_{MI} = C_p^{\frac{1}{\xi_s+\frac{5}{3}}}C_{ZP}^{-\frac{1}{\xi_s+\frac{5}{3}}}\varepsilon_u^{-\frac{1}{2}}\varepsilon_s^{\frac{3}{2}} \qquad (11b)$$

Since the MFD subrange is coincident in both $E_u$ and $E_s$, the two perspectives should give the same $k_{MI}$. This gives $C_{ZK} = C_K C_p^{-\left(\xi_u+\frac{5}{3}\right)/\left(\xi_s+\frac{5}{3}\right)}C_{ZP}^{\left(\xi_u+\frac{5}{3}\right)/\left(\xi_s+\frac{5}{3}\right)}$. According to Quad-cascade process model [14], $\xi_u$ and $\xi_s$ are both multivalue functions of $\beta$. Therefore, $C_{ZK}$ and $C_{ZP}$ can be different regarding the scaling exponents $\xi_u$ and $\xi_s$ of the specific MFD sub-branch (Table 1).

In Eqs. (11a, b), $k_{MI}$ is determined by $\varepsilon_u^{-1/2}$ and $\varepsilon_s^{3/2}$, showing a compromise of the two cascades of turbulent kinetic energy and scalar variance. When $\beta > 2/3$, the turbulent kinetic energy injected in the MFD subrange is transported inversely to inertial subrange. A higher $\varepsilon_u$ denotes how small wavenumber the turbulent kinetic energy in the MFD subrange can be extended backwardly. In contrast, the scalar variance experiences direct cascade from low to high wavenumber. A higher $\varepsilon_s$ denotes how large wavenumber the scalar variance in the MFD subrange can be extended forwardly. These two mechanisms give that $k_{MI}$ decreases with $\varepsilon_u$ but increases with $\varepsilon_s$.

$k_{MI}$ is not always distinguishable. Quad-cascade process model [14] concludes four sub-branches of subranges, regarding whether $\Pi_u$ and $\Pi_s$ are constant or not. The inertial subrange is one of the sub-branches. In this case, there is no $k_{MI}$ since only one inertial subrange is present. For this case, $C_{ZK} = C_K$ and $C_{ZP} = C_p$. While for the Constant-$\Pi_u$ subrange, $E_u \sim k^{-5/3}$ and $E_s \sim k^{-7/3}$. We cannot distinguish the MFD subrange with the inertial subrange from $E_u$, but can distinguish them from $E_s$. For this case, $k_{MI} = C_p^{-3/2}C_{ZP}^{3/2}\varepsilon_u^{-1/2}\varepsilon_s^{3/2}$, with $C_{ZK} = C_K$ and $C_{ZP}$ to be determined.

(2) The higher bound of the MFD subrange ($k_{MU}$)

The intersection wavenumbers between the MFD subrange and the dissipation subrange can be analyzed dimensionally as

$$k_{MD}u_k^2 = k_{MD}^{\beta} M s_k = c_u k_{MD}^{\gamma} u_k \tag{12a}$$

$$u_k k_{SD} = c_s k_{SD}^{\alpha} \tag{12b}$$

where $k_{MD}$ and $k_{SD}$ are the characteristic wavenumbers separating the MFD subrange and the dissipation subrange in turbulent kinetic energy spectra and scalar spectra respectively. Beyond $k_{MD}$ and $k_{SD}$, the fluxes of turbulence kinetic energy and scalar variance are overwhelmed by viscous and scalar diffusion respectively.

Substituting Eqs. (9a, b) into Eqs. (12a, b) yields

$$k_{MD} = \left(\varepsilon_u^{\frac{3+\xi_u}{4}} \varepsilon_s^{-\frac{5+3\xi_u}{4}} c_u^{-1}\right)^{\frac{1}{\gamma-\frac{(\xi_u+3)}{2}}} = \left(\frac{c_u}{M}\varepsilon_u^{\frac{2+\xi_u-\xi_s}{4}} \varepsilon_s^{\frac{3(\xi_s-\xi_u)-2}{4}}\right)^{\frac{1}{\beta-\gamma+\frac{\xi_s-\xi_u}{2}}} \tag{13a}$$

$$k_{SD} = \left(\varepsilon_u^{\frac{3+\xi_u}{4}} \varepsilon_s^{-\frac{5+3\xi_u}{4}} c_s^{-1}\right)^{\frac{1}{\alpha-\frac{\xi_u+3}{2}}} \tag{13b}$$

Given $M = |\boldsymbol{M}|$, $c_u$, $c_s$, $\gamma$, $\beta$, $\alpha$, and the scaling exponents $\xi_u$ and $\xi_s$ of the specific MFD sub-branch, $k_{MD}$ and $k_{SD}$ can be determined. Since $k_{MD}$ and $k_{SD}$ are determined separated from Eqs. (12a, b) independently, these two wavenumbers may differ significantly depending on $Sc_Z$. Therefore, the actual higher bound wavenumber of MFD subrange is defined as $k_{MU} = \min(k_{MD}, k_{SD})$. Beyond $k_{MU}$, novel spectral subranges emerge as will be discussed in section 6.

From $k_{MD}$ in Eq. (13a), it can be inferred that $\varepsilon_u$ is not independent, but relies on $M$, $c_u$ and $\varepsilon_s$ as

$$\varepsilon_u = M^{a3} c_u^{b3} \varepsilon_s^{c3} \tag{14}$$

where $a3 = -\frac{4(2\gamma-\xi_u-3)}{(3+\xi_u)(2\beta-2\gamma+\xi_s-\xi_u)-(2+\xi_u-\xi_s)(2\gamma-\xi_u-3)}$, $b3 = \frac{4(2\beta+\xi_s-2\xi_u-3)}{(3+\xi_u)(2\beta-2\gamma+\xi_s-\xi_u)-(2+\xi_u-\xi_s)(2\gamma-\xi_u-3)}$ and $c3 = \frac{(3\xi_s-3\xi_u-2)(2\gamma-\xi_u-3)+(5+3\xi_u)(2\beta-2\gamma+\xi_s-\xi_u)}{(3+\xi_u)(2\beta-2\gamma+\xi_s-\xi_u)-(2+\xi_u-\xi_s)(2\gamma-\xi_u-3)}$. For ordinary diffusions of momentum and scalars where $\gamma = \alpha = 2$, we further get

$$k_{MD} = \left(\varepsilon_u^{\frac{3+\xi_u}{4}} \varepsilon_s^{-\frac{5+3\xi_u}{4}} c_u^{-1}\right)^{\frac{2}{1-\xi_u}} = \left(\frac{c_u}{M}\varepsilon_u^{\frac{2+\xi_u-\xi_s}{4}} \varepsilon_s^{\frac{3(\xi_s-\xi_u)-2}{4}}\right)^{\frac{1}{\frac{\xi_s-\xi_u}{2}+\beta-2}} \tag{15a}$$

$$k_{SD} = \left(\varepsilon_u^{\frac{3+\xi_u}{4}} \varepsilon_s^{-\frac{5+3\xi_u}{4}} c_s^{-1}\right)^{\frac{2}{1-\xi_u}} \tag{15b}$$

$$\varepsilon_u = \left[M^{2(\xi_u-1)} c_u^{2\xi_s-4\xi_u-6+4\beta} \varepsilon_s^{(3\beta\xi_u+5\beta-9\xi_u+4\xi_s-11)}\right]^{\frac{1}{\beta(\xi_u+3)+2\xi_s-3\xi_u-7}} \tag{15c}$$

When $\beta = 1$, four pairs of $\xi_u$ and $\xi_s$ can be predicted (Table 1) according to Quad-cascade process model [14]. In the case Constant-$\Pi_u$ subrange where $\xi_u = -5/3$ and $\xi_s = -7/3$, $k_{MD} = (\varepsilon_u c_u^{-3})^{1/(3\gamma-2)} = k_K$, $k_{SD} = (\varepsilon_u c_s^{-3})^{1/(3\alpha-2)} = k_S$ with $\varepsilon_u = M\varepsilon_s$. The definitions of $k_K$ (reciprocal of Kolmogorov scale) and $k_S$ (reciprocal of scalar dissipation scale) can be found in Part 1 [12]. While in the case Constant-$\Pi_s$ subrange where $\xi_u = -7/5$ and $\xi_s = -9/5$, $k_{MD} = (\varepsilon_u^2 \varepsilon_s^{-1} c_u^{-5})^{1/(5\gamma-4)}$, $k_{SD} = (\varepsilon_u^2 \varepsilon_s^{-1} c_s^{-5})^{1/(5\alpha-4)}$ with $\varepsilon_u = M\varepsilon_s$ as well. Therefore, the energy feeding rate (through $M$) and scalar dissipation rate $\varepsilon_s$ determine how the kinetic energy and scalar variance are delivered. Since $\varepsilon_u$ and $\varepsilon_s$ must be primarily attributed to the multiscale force, if the turbulent kinetic energy is generated by a transfer from mean flow field, Eqs. (12) and (13) are not applicable.

Table 1. In all the Quad-cascade sub-branches of electrokinetic turbulence [15,16], the influence of the order of fractional derivation in anomalous diffusion on the characteristic wavenumbers of velocity and scalar, where $\beta = 1$. $k_{ADMS} = (c_u/c_s)^{1/(\alpha-\gamma)}$ is the corresponding characteristic wavenumber that the anomalous diffusions of momentum and scalar experience the same diffusion time. Anomalous Schmidt number $Sc_z = k_0^{\gamma-\alpha}\, c_u/c_s$, where $k_0$ is the lowest wavenumber. If $\alpha \neq \gamma$, $Sc_z = (k_{ADMS}/k_0)^{\alpha-\gamma}$, or alternatively $k_{ADMS} = Sc_z^{1/(\alpha-\gamma)} k_0$.

| | Inertial subrange | Constant-$\Pi_u$ subrange | Constant-$\Pi_s$ subrange | Variable fluxes (VF) subrange |
|---|---|---|---|---|
| Scaling exponents | $\xi_u = \xi_s = -\frac{5}{3}$ | $\xi_u = -\frac{5}{3}$<br>$\xi_s = -\frac{7}{3}$ | $\xi_u = -\frac{7}{5}$<br>$\xi_s = -\frac{9}{5}$ | $\xi_u = -2$<br>$\xi_s = -3$ |
| $C_{ZK}$ | $C_K$ | $C_K$ | $C_K C_p^2 C_{ZP}^{-2}$ | $C_K C_p^{-1/4} C_{ZP}^{1/4}$ |
| $k_{MI}$ | Not available | $C_p^{-\frac{3}{2}} C_{ZP}^{\frac{3}{2}} \varepsilon_u^{-\frac{1}{2}} \varepsilon_s^{\frac{3}{2}}$<br>$= C_p^{-\frac{3}{2}} C_{ZP}^{\frac{3}{2}} M^{-\frac{1}{2}} \varepsilon_s$ | $C_K^{\frac{15}{4}} C_{ZK}^{-\frac{15}{4}} \varepsilon_u^{-\frac{1}{2}} \varepsilon_s^{\frac{3}{2}}$<br>$= C_K^{\frac{15}{4}} C_{ZK}^{-\frac{15}{4}} M^{-\frac{1}{2}} \varepsilon_s$ | $C_K^{-3} C_{ZK}^{-3} \varepsilon_u^{-\frac{1}{2}} \varepsilon_s^{\frac{3}{2}}$<br>$= C_K^{-3} C_{ZK}^{-3} M^{-\frac{1}{2}} \varepsilon_s$ |
| $k_{MD}$ | $(\varepsilon_u c_u^{-3})^{\frac{1}{(3\gamma-2)}}$ | $(\varepsilon_u c_u^{-3})^{\frac{1}{(3\gamma-2)}}$ | $(\varepsilon_u^2 \varepsilon_s^{-1} c_u^{-5})^{\frac{1}{(5\gamma-4)}}$ | $(\varepsilon_u \varepsilon_s c_u^{-4})^{\frac{1}{4\gamma-2}}$ |
| $k_{SD}$ | $(\varepsilon_u c_s^{-3})^{\frac{1}{(3\alpha-2)}}$<br>$= k_{MD}^{\frac{3\gamma-2}{3\alpha-2}} k_0^{\frac{-3(\gamma-\alpha)}{3\alpha-2}} Sc_z^{\frac{3}{3\alpha-2}}$<br>$= k_{MD}^{\frac{3\gamma-2}{3\alpha-2}} \Big/ k_{ADMS}^{\frac{3(\gamma-\alpha)}{3\alpha-2}}$ | $(\varepsilon_u c_s^{-3})^{\frac{1}{(3\alpha-2)}}$<br>$= k_{MD}^{\frac{3\gamma-2}{3\alpha-2}} k_0^{\frac{-3(\gamma-\alpha)}{3\alpha-2}} Sc_z^{\frac{3}{3\alpha-2}}$<br>$= k_{MD}^{\frac{3\gamma-2}{3\alpha-2}} \Big/ k_{ADMS}^{\frac{3(\gamma-\alpha)}{3\alpha-2}}$ | $(\varepsilon_u^2 \varepsilon_s^{-1} c_s^{-5})^{\frac{1}{(5\alpha-4)}}$<br>$= k_{MD}^{\frac{(5\gamma-4)}{5\alpha-4}} k_0^{\frac{-5(\gamma-\alpha)}{5\alpha-4}} Sc_z^{\frac{5}{5\alpha-4}}$<br>$= k_{MD}^{\frac{(5\gamma-4)}{5\alpha-4}} \Big/ k_{ADMS}^{\frac{5(\gamma-\alpha)}{5\alpha-4}}$ | $(\varepsilon_u \varepsilon_s c_s^{-4})^{\frac{1}{4\alpha-2}}$<br>$= k_{MD}^{\frac{2\gamma-1}{2\alpha-1}} k_0^{\frac{-2(\gamma-\alpha)}{2\alpha-1}} Sc_z^{\frac{2}{2\alpha-1}}$<br>$= k_{MD}^{\frac{2\gamma-1}{2\alpha-1}} \Big/ k_{ADMS}^{\frac{2(\gamma-\alpha)}{2\alpha-1}}$ |
| $\varepsilon_u$ | $M^{\frac{6\gamma-4}{5\gamma-4}} c_u^{-\frac{2}{5\gamma-4}} \varepsilon_s^{\frac{3\gamma-2}{5\gamma-4}}$ | $M\varepsilon_s$ | $M\varepsilon_s$ | $M\varepsilon_s$ |

One important relationship from Eqs. (8) and (9) is $\xi_u = (\xi_s - 1)/2$. This relationship establishes in the Constant-$\Pi_u$ subrange, the Constant-$\Pi_s$ subrange and the VF subrange, except the inertial subrange. This is because $\xi_u = \xi_s = -5/3$ in the inertial subrange is a special solution of Eq. (7), which can be reached without Eq. (8).

## 4. Anomalous diffusions and the characteristic scales of EK turbulence

In this subsection and below, we focus on electrokinetic turbulence where $\beta = 1$. As studied by Golestanian [17], the electrolyte can experience several anomalous diffusions under electric field in different scale range (Table 2). At a short time, say $t < \tau_{ba}$, the electrolyte experiences a ballistic transport, where super-diffusion with $\alpha = 1$ dominates electrolyte transport. This is corresponding to a high wavenumber region $k > k_{ba}$. When $k_{SD} \geq k_{ba}$, it can be guaranteed that the entire scalar dissipation subrange is within the ballistic region where $\alpha = 1$. However, if $k_{ax} \ll k_{SD} \ll k_{ba}$, a second anomalous diffusion region becomes dominant where $\alpha = 4/3$. It leads to a faster scalar transport at medium time. If $k_{SD} \ll k_{ax}$, a long-time anomalous diffusion becomes dominant with $\alpha = 3/2$. This consecutive anomalous diffusion feature makes the analysis on the structure of turbulence ultracomplex, particular in the diagram of Quad-cascade process model [14-16]. Therefore, to reveal how the anomalous diffusions affecting the cascades and the characteristic wavenumbers in the momentum-scalar coupling turbulence, we only consider each single anomalous diffusion once.

Table 2. Order of fractional derivation in anomalous diffusion of electrokinetic flow where $\beta = 1$. Summarized according to Golestanian [17].

| Diffusion Phases | MSD | $\alpha$ | Definitions |
|---|---|---|---|
| Short-time ballistic region ($0 < t < \tau_{ba}$) | $\Delta L^2 \sim t^{\frac{2}{\alpha}}$ | 1 | • $\tau_{ba} = d_m^2 D^{-1}$ is the diffusion time scale separating the short-time ballistic region and the medium-time anomalous region. The corresponding $\Delta L = (c_{s1}\tau_{ba})^{\frac{1}{\alpha}} = c_{s1} d_m^2 D^{-1}$ and wavenumber $k_{ba} = 2\pi c_{s1}^{-1} d_m^{-2} D$. Here, $D$ is the diffusivity of ordinary diffusion, $d_m = \max[d_p, \lambda]$ is the larger one between the tracer size ($d_p$) and Debye length $\lambda = \sqrt{\epsilon k_B T / 2 S_d C_0 Q^2}$ , $C_0$ is the mean ion concentration, $\epsilon$ is the electric permittivity, $k_B$ is Boltzmann constant, $T$ is temperature in Kelvin, $S_d = 2\pi^{\frac{d}{2}} / \Gamma\left(\frac{d}{2}\right)$ is the d-dimensional spherical surface area factor which is $4\pi$ in 3D space. According to Einstein-Stokes equation that $D = k_B T / 3\pi\eta d_p$, we further have $k_{ba} = 2 k_B T / 3\eta d_m^2 d_p c_{s1}$ , where $\eta = \rho c_u$ is the solvent viscosity. |
| Medium-time anomalous region ($\tau_{ba} < t < \tau_{ax}$) | | 4/3 | |
| Long-time anomalous region ($t > \tau_{ax}$) | | 3/2 | • $\tau_{ax} = C_0^2 D^3 \tau_e^4$ is the time scale separating the medium-time ballistic region and the long-time anomalous region. The corresponding mean-square displacement $\Delta L^2 = (c_{s1}\tau_{ba})^2 + [c_{s2}(\tau_{ax} - \tau_{ba})]^{\frac{3}{2}}$ and wavenumber $k_{ax} = 2\pi / \sqrt{(c_{s1} d_m^2 D^{-1})^2 + [c_{s2}(C_0^2 D^3 \tau_e^4 - d_m^2 D^{-1})]^{\frac{3}{2}}}$ , where $\tau_e = (\epsilon E^2 / 2 S_d \eta)^{-1}$ is the characteristic time scale corresponding to the shear strain rate due to Maxwell stress associated with the electric field, $E$ is the electric field. |

As examples, we plot Fig. 2 to demonstrate how the characteristic wavenumbers (i.e. $k_{ba}$ and $k_{ax}$) are determined by the control parameters. Since the anomalous diffusion in EK flow is induced under electric field, the influence of $E$ is investigated first, as shown in Fig. 2(a). Under the conditions, $k_{ba} = 1.3 \times 10^{14}$ m$^{-1}$ which is irrelevant to $E$, while $k_{ax}$ increases from $3.8 \times 10^{-18}$ m$^{-1}$ to 3.8 m$^{-1}$. In the region where this physical model is applicable, i.e. orders of $10^{-3}$ m$^{-1}$ (large scale circulation, e.g. in atmosphere) to $2\pi \times 10^9$ m$^{-1}$ (small scale for continuum), both the medium-time anomalous and long-time anomalous diffusions of ions can be present. Another important parameter is $D$. Fig. 2(b) shows $k_{ba}$ increases with $D$, while $k_{ax}$ decreases rapidly. In the lower limit of $D$ in the current investigation, the region where this physical model is applicable covers all the three anomalous diffusion regions of ions. Therefore, all the three anomalous diffusions are discussed in the following sections.

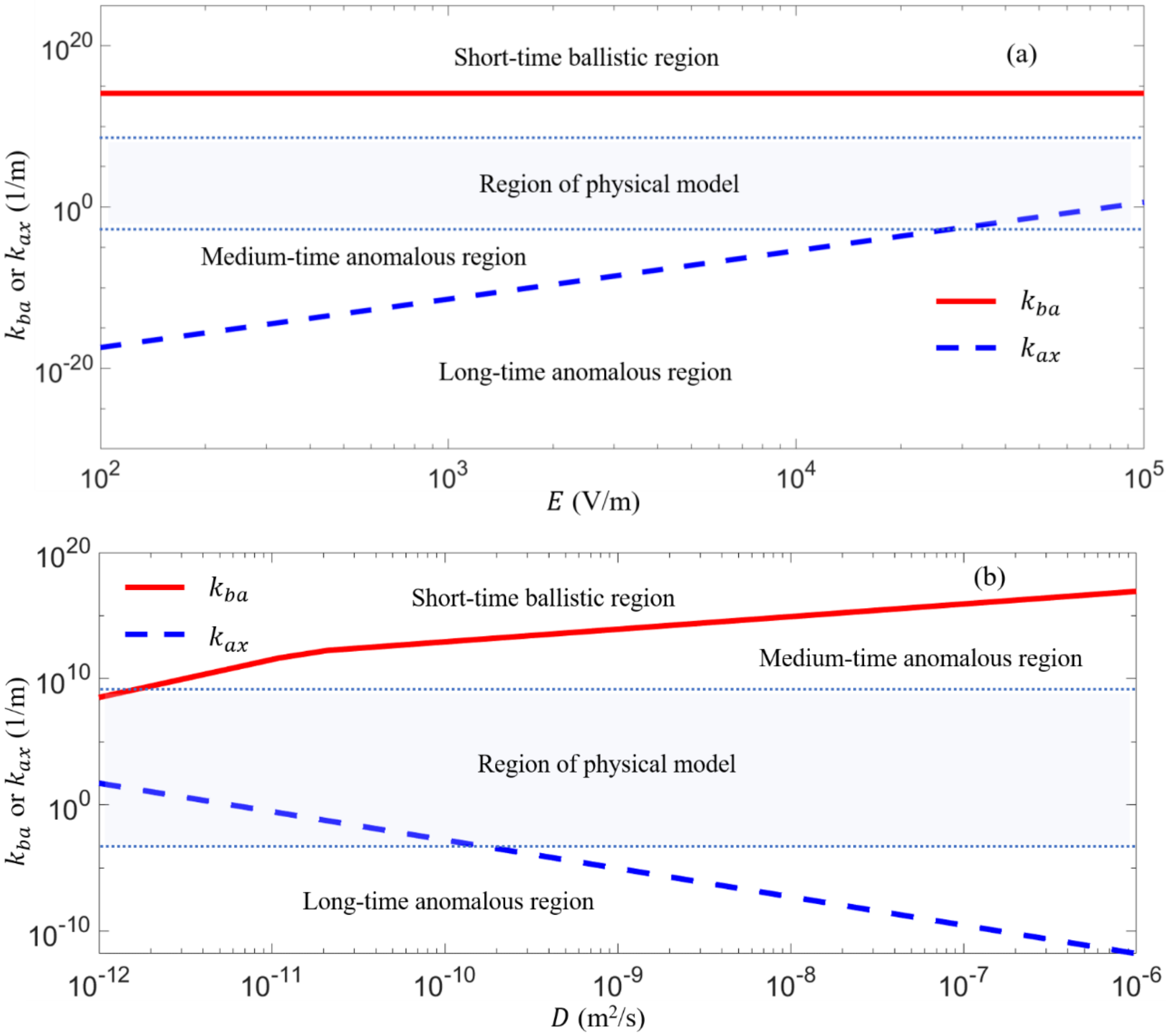

Fig. 2 $k_{ba}$, $k_{ax}$ determined by the control parameters. Here, single-atom ions are considered as examples, where $d_p = 2.9 \times 10^{-10}$ m, $\epsilon = \epsilon_0 \epsilon_r$, $\epsilon_0 = 8.854 \times 10^{-12}$ F/m, $\epsilon_r = 80.1$ for water, $K_B = 1.380 \times 10^{-23}$ J/K is Boltzmann constant, $T = 298$ K, $Q = 1.602 \times 10^{-19}$ C, $NA$ is Avogadro constant, $C_M = 10^{-2}$ mol/m$^3$ is ionic molar concentration, $C_0 = C_M \times NA$. (a) $c_{s1} = 10^{-7}$ m/s, $c_{s2} = 10^{-7}$ m$^{4/3}$/s, $D = 1.5 \times 10^{-9}$ m$^2$/s, $c_u = 10^{-6}$ m$^2$/s, $\rho = 10^3$ kg/m$^3$, and (b) $E = 10^5$ V/m, $\lambda = 2.741 \times 10^{-8}$ m, $c_{s1} = 10^{-7}$ m/s, $c_{s2} = 10^{-7}$ m$^{4/3}$/s, $c_u = 10^{-6}$ m$^2$/s.

## 5. Characteristic scales in EK turbulence where $\beta = 1$

The characteristic wavenumbers are theoretically summarized in Table 3. It can be seen, $k_{MD}$ is primarily determined by which subrange is interplayed with dissipation subrange. It is not superficially affected by the anomalous diffusion of scalar. However, since $\varepsilon_u$ is a function of $\varepsilon_s$ which is intrinsically and implicitly determined by $\alpha$, $k_{MD}$ is also implicityly related to $\alpha$ as well.

In contrast, the relationship between $k_{SD}$ and $\alpha$ is highly visible. For instance, when $\gamma = 2$ and $\alpha = 1$, $k_{SD} = \varepsilon_u^2 \varepsilon_s^{-1} c_s^{-5} = k_{MD}(k_{MD}/k_0)^5 Sc_z^5$ if the Constant-$\Pi_s$ subrange is interconnected with the dissipation subrange. When $\alpha = 4/3$, the situation becomes $k_{SD} = \varepsilon_u^{3/4} \varepsilon_s^{-3/8} c_s^{-15/8} = k_{MD}(k_{MD}/k_0)^{5/4} Sc_z^{15/8}$. Regarding $k_{MD}/k_0 \gg 1$, the relationship between $k_{MD}$ and $k_{SD}$ is further determined by $Sc_z$, which is controlled by $\alpha$ and $\gamma$ as well.

Table 3. In all the Quad-cascade sub-branches of electrokinetic turbulence [15,16], the influence of the order of fractional derivation in anomalous diffusion on the characteristic wavenumbers of velocity and scalar, where $\beta = 1$ and $\gamma = 2$.

| MFD subrange | | Inertial subrange | Constant-$\Pi_u$ subrange | Constant-$\Pi_s$ subrange | VF subrange |
|---|---|---|---|---|---|
| $k_{MD}$ | | $\varepsilon_u^{\frac{1}{4}} c_u^{-\frac{3}{4}}$ | $\varepsilon_u^{\frac{1}{4}} c_u^{-\frac{3}{4}}$ | $\varepsilon_u^{\frac{1}{3}} \varepsilon_s^{-\frac{1}{6}} c_u^{-\frac{5}{6}}$ | $\varepsilon_u^{\frac{1}{6}} \varepsilon_s^{\frac{1}{6}} c_u^{-\frac{2}{3}}$ |
| $k_{SD}$ | $\alpha = 1$ | $\varepsilon_u c_s^{-3} = k_{MD}^4 k_0^{-3} Sc_z^3$ | $\varepsilon_u c_s^{-3} = k_{MD}^4 k_0^{-3} Sc_z^3$ | $\varepsilon_u^2 \varepsilon_s^{-1} c_s^{-5}$ $= k_{MD}^6 k_0^{-5} Sc_z^5$ | $\varepsilon_u^{\frac{1}{2}} \varepsilon_s^{\frac{1}{2}} c_s^{-2}$ $= k_{MD}^3 k_0^{-2} Sc_z^2$ |
| | $\alpha = \frac{4}{3}$ | $\varepsilon_u^{\frac{1}{2}} c_s^{-\frac{3}{2}} = k_{MD}^2 k_0^{-1} Sc_z^{\frac{3}{2}}$ | $\varepsilon_u^{\frac{1}{2}} c_s^{-\frac{3}{2}} = k_{MD}^2 k_0^{-1} Sc_z^{\frac{3}{2}}$ | $\varepsilon_u^{\frac{3}{4}} \varepsilon_s^{-\frac{3}{8}} c_s^{-\frac{15}{8}}$ $= k_{MD}^{\frac{9}{4}} k_0^{-\frac{5}{4}} Sc_z^{\frac{15}{8}}$ | $(\varepsilon_u \varepsilon_s c_s^{-4})^{\frac{3}{10}}$ $= k_{MD}^{\frac{9}{5}} k_0^{-\frac{4}{5}} Sc_z^{\frac{6}{5}}$ |
| | $\alpha = \frac{3}{2}$ | $\varepsilon_u^{\frac{2}{5}} c_s^{-\frac{6}{5}} = k_{MD}^{\frac{8}{5}} k_0^{\frac{-3}{5}} Sc_z^{\frac{6}{5}}$ | $\varepsilon_u^{\frac{2}{5}} c_s^{-\frac{6}{5}} = k_{MD}^{\frac{8}{5}} k_0^{\frac{-3}{5}} Sc_z^{\frac{6}{5}}$ | $\varepsilon_u^{\frac{4}{7}} \varepsilon_s^{-\frac{2}{7}} c_s^{-\frac{10}{7}}$ $= k_{MD}^{\frac{12}{7}} k_0^{-\frac{5}{7}} Sc_z^{\frac{10}{7}}$ | $\varepsilon_u^{\frac{1}{4}} \varepsilon_s^{\frac{1}{4}} c_s^{-1}$ $= k_{MD}^{\frac{3}{2}} k_0^{-\frac{1}{2}} Sc_z$ |
| $\varepsilon_u$ | | $M^{\frac{4}{3}} c_u^{-\frac{1}{3}} \varepsilon_s^{\frac{2}{3}}$ | $M\varepsilon_s$ | $M\varepsilon_s$ | $M\varepsilon_s$ |

The variations of $k_{MD}$ and $k_{SD}$ with multiple parameters have been plotted in Fig. 3. From Table 3, it can be seen both $k_{MD}$ and $k_{SD}$ are, at first glance, determined by $\varepsilon_u$ which is dominated by $\varepsilon_s$. Fig. 3(a) shows how $k_{MD}$ and $k_{SD}$ vary with $\varepsilon_s$. For simplicity, we assume $\gamma = 2$, i.e. ordinary momentum diffusion. In all the Quad-cascade subranges, $k_{MD}$ and $k_{SD}$ all increase with $\varepsilon_s$ as power laws. Particularly, their scaling exponents varying with $\varepsilon_s$ are highly dependent on the Quad-cascade sub-branches and $\alpha$.

In all the Quad-cascade sub-branches, $k_{MD}$ lies below the upper bound wavenumber ($k_{UBPM}$) of the physical model. It should therefore be regarded as an effective and important characteristic scale of velocity structure in EK turbulence. In contrast, $k_{SD}$ spans a wide wavenumber range. For instance, in the inertial subrange where $\Pi_u$ and $\Pi_s$ are both constant (up-right of Fig. 3(a)), $k_{SD}$ can be up to $2 \times 10^{20}$ m$^{-1}$ at $\alpha = 1$ which is much higher than the $k_{UBPM}$. Thus, $k_{SD}$ in this case is too high to be realized in a real EK flow. However, at $\alpha = 2$, $k_{SD}$ becomes overall below $k_{UBPM}$. In this subrange of EK turbulence, $k_{SD}$ at $\alpha = 1$, 4/3 and 3/2 are all over $k_{UBPM}$ in the considered range of $\varepsilon_s$. Therefore, it is unnecessary to discuss $k_{SD}$ in this subrange of EK turbulence, except $\varepsilon_s$ is too small (very dilute ions).

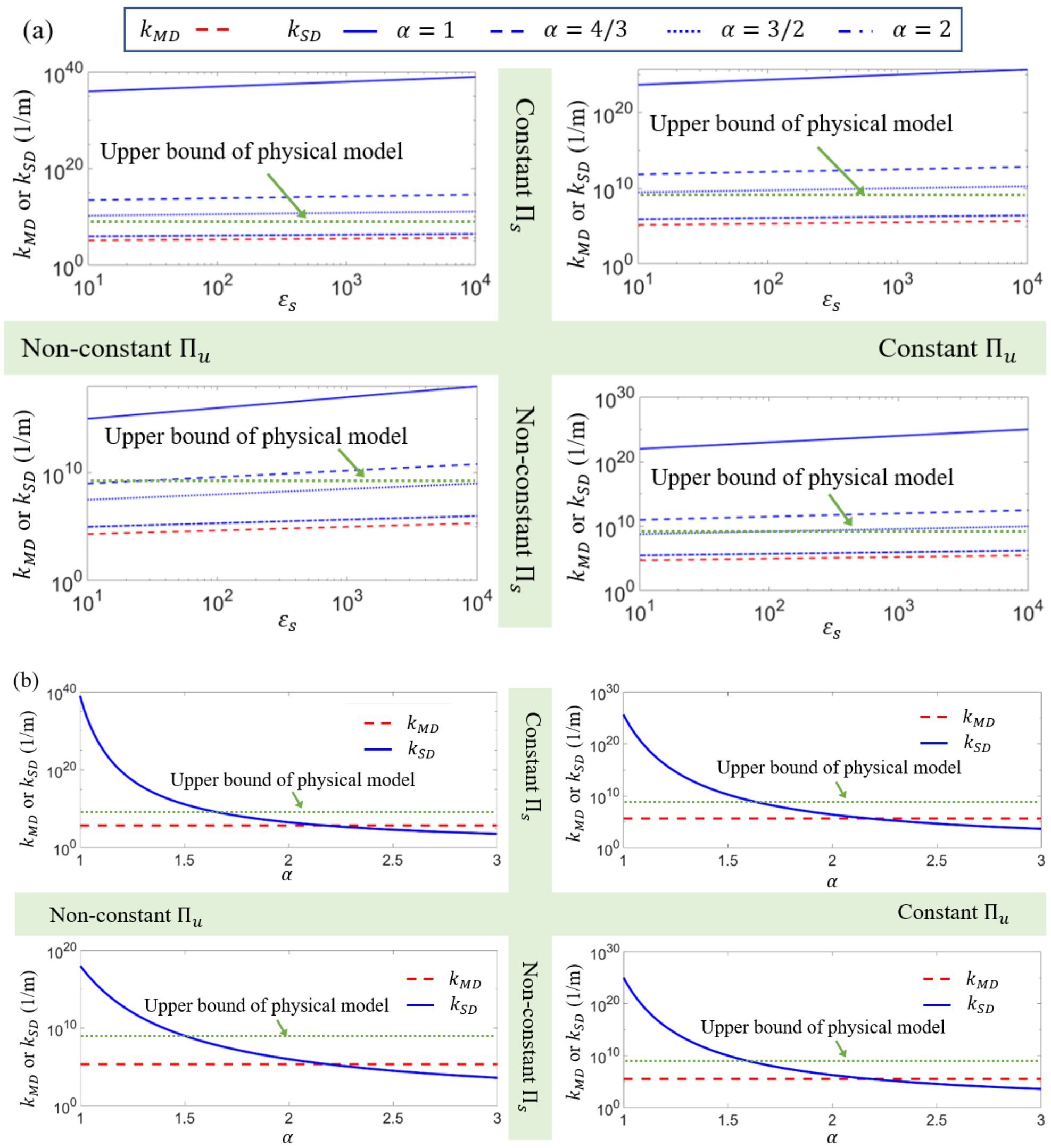

Fig. 3 Relationship among $k_{MD}$ and $k_{SD}$ under different $\varepsilon_u$, $\varepsilon_s$, $c_u$ and $c_s$ for the four Quad-cascade sub-branches. (a) $k_{MD}$ and $k_{SD}$ as functions of $\varepsilon_s$, where $c_s = 10^{-7}$ m/s, $c_u = 10^{-6}$ m$^2$/s, $\rho = 10^3$ kg/m$^3$, $M = 1$, $\gamma = 2$. (b) $k_{MD}$ and $k_{SD}$ as functions of $\alpha$, where $c_s = 10^{-7}$ m/s, $c_u = 10^{-6}$ m$^2$/s, $\rho = 10^3$ kg/m$^3$, $\varepsilon_s = 10^4$ 1/s, $M = 1$, $\gamma = 2$. The GREEN dashed line indicates the upper bound of the physical model $k_{UBPM}$.

While in the VF subrange where $\Pi_u$ and $\Pi_s$ are both non-constant (bottom-left of Fig. 3(a)), $k_{SD}$ can be below the $k_{UBPM}$ at a lower $\varepsilon_s$, when $\alpha = 4/3$. In this case, it is necessary to take into account the $k_{SD}$ regarding the medium- and long-time anomalous diffusions of ions.

Fig. 3(b) further demonstrates how $\alpha$ dominates $k_{SD}$. Since $k_0$ in this study is assumed to be 10 m$^{-1}$, in the range $k > k_0$, a superdiffusion with $\alpha < 2$ leads to a slow diffusion as have been introduced in Part 1 [12]. The smaller the $\alpha$, the slower the diffusion is, accordingly, a larger $k_{SD}$ can be inferred for all the Quad-cascade sub-branches. The difference is, in the VF subrange, $k_{SD}$ can be lower than the $k_{UBPM}$ at $\alpha > 1.52$ (this number is not universal, but depends on specific conditions), while in the Constant-$\Pi_s$ subrange, this is realized at $\alpha > 1.69$. Therefore, all the Quad-cascade sub-branches in EK turbulence are established only in the range $k \leq \min(k_{UBPM}, k_{SD})$. As $\alpha$ is further increased to over 2, the scalar experiences subdiffusion. In the range $k > k_0$, the diffusion of scalar could be even faster than momentum with $Sc_z < 1$, leading to $k_{SD} < k_{MD}$.

The influence of anomalous momentum diffusion on the characteristic wavenumber in EK turbulence is also theoretically elucidated in Table 4 as reference to the readers. All these results demonstrate $\gamma$, $\alpha$ and the related $Sc_z$ control the structures and their characteristic wavenumbers in EK turbulence. In the following section, the velocity and scalar spectra additional to the MFD subranges will be discussed upon $Sc_z$.

Table 4. Influence of the order of fractional derivation in anomalous diffusion of electrokinetic turbulence on the characteristic wavenumbers of velocity and scalar, where $\beta = 1$ and $\alpha = 2$.

| | | Inertial subrange | Constant-$\Pi_u$ subrange | Constant-$\Pi_s$ subrange | VF subrange |
|---|---|---|---|---|---|
| $k_{MD}$ | $\gamma = \frac{3}{2}$ | $\varepsilon_u^{\frac{2}{5}} c_u^{-\frac{6}{5}}$ | $\varepsilon_u^{\frac{2}{5}} c_u^{-\frac{6}{5}}$ | $\varepsilon_u^{\frac{4}{7}} \varepsilon_s^{-\frac{2}{7}} c_u^{-\frac{10}{7}}$ | $\varepsilon_u^{\frac{1}{4}} \varepsilon_s^{\frac{1}{4}} c_u^{-1}$ |
| | $\gamma = 2$ | $\varepsilon_u^{\frac{1}{4}} c_u^{-\frac{3}{4}}$ | $\varepsilon_u^{\frac{1}{4}} c_u^{-\frac{3}{4}}$ | $\varepsilon_u^{\frac{1}{3}} \varepsilon_s^{-\frac{1}{6}} c_u^{-\frac{5}{6}}$ | $\varepsilon_u^{\frac{1}{6}} \varepsilon_s^{\frac{1}{6}} c_u^{-\frac{2}{3}}$ |
| | $\gamma = \frac{5}{2}$ | $\varepsilon_u^{\frac{2}{11}} c_u^{-\frac{6}{11}}$ | $\varepsilon_u^{\frac{2}{11}} c_u^{-\frac{6}{11}}$ | $\varepsilon_u^{\frac{4}{17}} \varepsilon_s^{-\frac{2}{17}} c_u^{-\frac{10}{17}}$ | $\varepsilon_u^{\frac{1}{8}} \varepsilon_s^{\frac{1}{8}} c_u^{-\frac{1}{2}}$ |
| $k_{SD}$ | $\gamma = \frac{3}{2}$ | $\varepsilon_u^{\frac{1}{4}} c_s^{-\frac{3}{4}} = k_{MD}^{\frac{5}{8}} k_0^{\frac{3}{8}} Sc_z^{\frac{3}{4}}$ | $\varepsilon_u^{\frac{1}{4}} c_s^{-\frac{3}{4}} = k_{MD}^{\frac{5}{8}} k_0^{\frac{3}{8}} Sc_z^{\frac{3}{4}}$ | $\varepsilon_u^{\frac{1}{3}} \varepsilon_s^{-\frac{1}{6}} c_s^{-\frac{5}{6}}$ $= k_{MD}^{\frac{7}{12}} k_0^{\frac{5}{12}} Sc_z^{\frac{5}{6}}$ | $\varepsilon_u^{\frac{1}{6}} \varepsilon_s^{\frac{1}{6}} c_s^{-\frac{2}{3}}$ $= k_{MD}^{\frac{2}{3}} k_0^{\frac{1}{3}} Sc_z^{\frac{2}{3}}$ |
| | $\gamma = 2$ | $\varepsilon_u^{\frac{1}{4}} c_s^{-\frac{3}{4}} = k_{MD} Sc_z^{\frac{3}{4}}$ | $\varepsilon_u^{\frac{1}{4}} c_s^{-\frac{3}{4}} = k_{MD} Sc_z^{\frac{3}{4}}$ | $\varepsilon_u^{\frac{1}{3}} \varepsilon_s^{-\frac{1}{6}} c_s^{-\frac{5}{6}} = k_{MD} Sc_z^{\frac{5}{6}}$ | $\varepsilon_u^{\frac{1}{6}} \varepsilon_s^{\frac{1}{6}} c_s^{-\frac{2}{3}} = k_{MD} Sc_z^{\frac{2}{3}}$ |
| | $\gamma = \frac{5}{2}$ | $\varepsilon_u^{\frac{1}{4}} c_s^{-\frac{3}{4}} = k_{MD}^{\frac{11}{8}} k_0^{-\frac{3}{8}} Sc_z^{\frac{3}{4}}$ | $\varepsilon_u^{\frac{1}{4}} c_s^{-\frac{3}{4}} = k_{MD}^{\frac{11}{8}} k_0^{-\frac{3}{8}} Sc_z^{\frac{3}{4}}$ | $\varepsilon_u^{\frac{1}{3}} \varepsilon_s^{-\frac{1}{6}} c_s^{-\frac{5}{6}}$ $= k_{MD}^{\frac{17}{12}} k_0^{-\frac{5}{12}} Sc_z^{\frac{5}{6}}$ | $\varepsilon_u^{\frac{1}{6}} \varepsilon_s^{\frac{1}{6}} c_s^{-\frac{2}{3}}$ $= k_{MD}^{\frac{4}{3}} k_0^{-\frac{1}{3}} Sc_z^{\frac{2}{3}}$ |
| $\varepsilon_u$ | $\gamma = \frac{3}{2}$ | $M^{\frac{10}{7}} c_u^{-\frac{4}{7}} \varepsilon_s^{\frac{5}{7}}$ | $M\varepsilon_s$ | $M\varepsilon_s$ | $M\varepsilon_s$ |
| | $\gamma = 2$ | $M^{\frac{4}{3}} c_u^{-\frac{1}{3}} \varepsilon_s^{\frac{2}{3}}$ | $M\varepsilon_s$ | $M\varepsilon_s$ | $M\varepsilon_s$ |
| | $\gamma = \frac{5}{2}$ | $M^{\frac{22}{17}} c_u^{-\frac{4}{17}} \varepsilon_s^{\frac{11}{17}}$ | $M\varepsilon_s$ | $M\varepsilon_s$ | $M\varepsilon_s$ |

## 6. Velocity and scalar spectra beyond MFD subrange

From Tables (1, 3, 4) above, the MFD subrange can significantly affect the velocity and scalar structures at high wavenumbers, depending on $Sc_z$. In constrast to the cases of no forcing or forcing with $\beta < 2/3$ in Part 1 [12], where the MFD subrange is not interconnected with the dissipation subrange, the MFD subrange directly interplays with the dissipation subrange at $\beta > 2/3$. In the former, $E_u$ of the inertial subranges is not coupled to $E_s$, showing different bandwidths, $k_K$ and $k_S$ according to $Sc_z$. However, in the latter, the MFD subrange relies on the synchronous coupling of momentum and scalar. The MFD subrange in both $E_u$ and $E_s$ locates in the same region, with the same lower ($k_{MI}$) and upper bounds ($k_{MU}$). This means, we cannot have $E_u$ terminates at $k_{MD}$ and $E_s$ terminates at $k_{SD}$, with $k_{MD}$ and $k_{SD}$ being unequal. Therefore, when the momentum diffusion and scalar diffusion have different dynamics, we can expect CFV and DFC subranges, which connect the MFD subrange and dissipation subranges, depending on $Sc_z$, $\gamma$ and $\alpha$. In this case, Eqs. (6a, b) are not established. Their complicated feedback loops have been diagramed in Fig. 4.

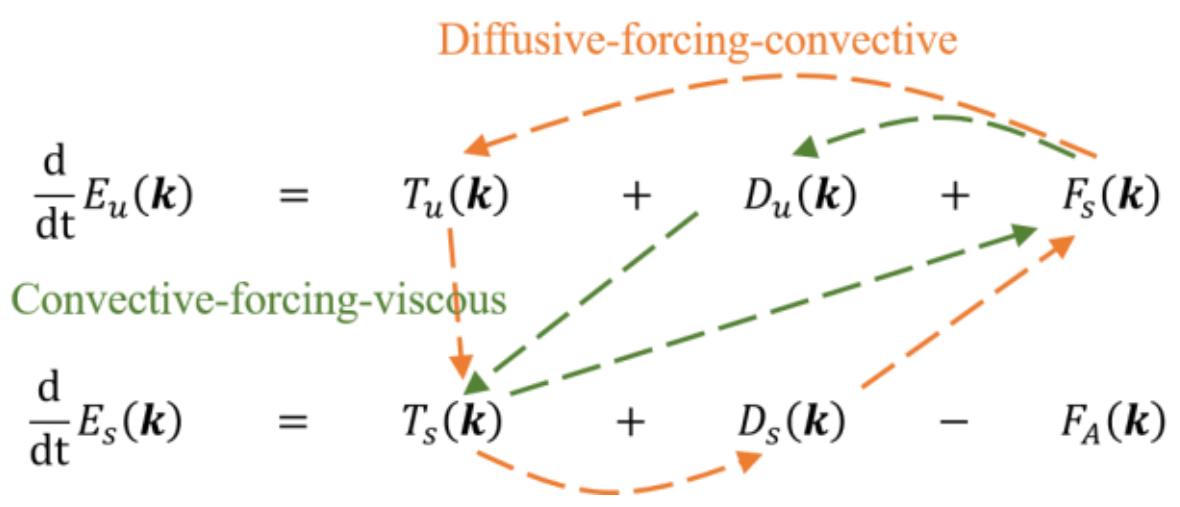

Fig. 4 Schematic of the interactions among the terms of governing equations about CFV and DFC subranges when $\beta > 2/3$.

*6.1 In the case $Sc_z \gg 1$*

When $Sc_z \gg 1$, the scalar diffusion is much weaker than momentum diffusion. The scalar structure can be delivered into a higher wavenumber, with $k_{SD} \gg k_{MD} = k_{MU}$. This is the CFV subrange where the influence of momentum diffusion becomes important, but the influence of scalar diffusion is negligible. It locates in $k_{MU} \ll k \ll k_{CFVD}$, with $k_{CFVD}$ being the upper bound of CFV subrange. The control equations in the subrange can be

$$\frac{\mathrm{d}}{\mathrm{d}k}\Pi_u(k) = F_s(k) + D_u(k) \tag{16$a$}$$

$$\frac{\mathrm{d}}{\mathrm{d}k}\Pi_s(k) = -F_A(k) \tag{16$b$}$$

Regarding $F_s(k) - \frac{M}{N}k^\beta F_A(k) = 0$,

$$\frac{\mathrm{d}}{\mathrm{d}k}\Pi_u(k) - D_u(k) + \frac{M}{N}k^\beta \frac{\mathrm{d}}{\mathrm{d}k}\Pi_s(k) = 0 \tag{17}$$

In the CFV subrange, although we still phenomenologically have $E_u \sim u_k^2/k$ and $E_s \sim s_k^2/k$, the fluxes relationship $\Pi_u(k) = ku_k^3$ and $\Pi_s(k) = ks_k^2 u_k$ may not establish. They strictly rely on the characteristic time scale $\tau$ in the form $\Pi_u(k) = E_u k/\tau$ and $\Pi_s(k) = E_s k/\tau$. Accordingly, Eq. (17) becomes

$$\frac{\mathrm{d}}{\mathrm{d}k}\left(\frac{E_u k}{\tau}\right) - 2c_u \mathrm{Re}\left(e^{i\frac{1}{2}\pi\gamma}\right)k^\gamma E_u + \frac{M}{N}k^\beta \frac{\mathrm{d}}{\mathrm{d}k}\left(\frac{E_s k}{\tau}\right) = 0 \tag{18}$$

In this subrange, the dominant portion of kinetic energy injected is directly dissipated, thus the forcing term and momentum diffusion term in Eq. (12a) could be phenomenologically balanced as

$$k^{\beta}Ms_k = c_u k^{\gamma} u_k \rightarrow u_k = k^{\beta-\gamma}Mc_u^{-1}s_k \tag{19}$$

This gives us a second relationship between velocity and scalar in $k$-space. Regarding different $\tau$, two possible scaling properties can be predicted, as shown in the following.

(1) If $\tau = 1/ku_k$ which is determined by convection time, the momentum convection is still important. $\Pi_u(k)$ and $\Pi_s(k)$ return to $ku_k^3$ and $ks_k^2u_k$ respectively. From Eq. (19) we have

$$\tau = k^{\gamma-\beta-\frac{3}{2}}M^{-1}c_u E_s^{-\frac{1}{2}} \tag{20}$$

and

$$E_u = k^{2\beta-2\gamma}M^2c_u^{-2}E_s \tag{21}$$

By substituting Eqs. (20, 21) into Eq. (18), simply we get

$$MNc_u^{-2}\frac{\mathrm{d}}{\mathrm{d}k}\left(k^{3\beta-3\gamma+5/2}E_s^{\frac{3}{2}}\right) + k^{\beta}\frac{\mathrm{d}}{\mathrm{d}k}\left(k^{-\gamma+\beta+\frac{5}{2}}E_s^{\frac{3}{2}}\right) - 2N\mathrm{Re}\left(e^{i\frac{1}{2}\pi\gamma}\right)k^{2\beta-\gamma}E_s = 0 \tag{22}$$

Eq. (22) is solved by Deepseek V4.1 and verified manually. For EK turbulence where $\beta = 1$, if $\gamma \neq 1/2$,

$$E_s(k) = k^{\frac{2\gamma-7}{3}}\left(MNc_u^{-2}k^{1-2\gamma}+1\right)^{-\frac{4(1-\gamma)}{3(1-2\gamma)}}\left[\frac{2N\mathrm{Re}\left(e^{i\frac{1}{2}\pi\gamma}\right)}{3}\int k^{-\frac{4+\gamma}{3}}\left(MNc_u^{-2}k^{1-2\gamma}+1\right)^{\frac{4\gamma-1}{3(1-2\gamma)}}dk + C_{CV}\right]^2 \tag{23}$$

where $C_{CV}$ is a constant to be determined. It has a dimension of $L^{4(\gamma-1)/3}$. Under a strong flow disturbance relative to the viscous effect, we can predict $MNc_u^{-2}k^{1-2\gamma} \gg 1$ in a broad range of $k$, with a characteristic wavenumber $k_{CV1} = (c_u^2/MN)^{1/(1-2\gamma)}$. In this investigation, we only consider $\gamma \geq 1$, thus $1-2\gamma < 0$. On one hand, this gives $MNc_u^{-2}k^{1-2\gamma} \gg 1$ in the range $k \ll k_{CV1}$. Eq. (23) gives

$$E_s(k) = k^{\frac{2\gamma-7}{3}}\left(MNc_u^{-2}k^{1-2\gamma}\right)^{-\frac{4(1-\gamma)}{3(1-2\gamma)}}\left[\frac{2N\left(MNc_u^{-2}\right)^{\frac{4\gamma-1}{3(1-2\gamma)}}\mathrm{Re}\left(e^{i\frac{1}{2}\pi\gamma}\right)}{3\gamma-2}k^{\gamma-\frac{2}{3}} + C_{CV}\right]^2 \tag{24}$$

Let $C_{CV1} = C_{CV}^2/\left[Nk_{CV1}^{-(\gamma+1)/3}\right]^2$ being a dimensionless constant, we further get

$$E_s(k) = C_{CV1}N^2k_{CV1}^{\frac{2}{3}-2\gamma}k^{2\gamma-\frac{11}{3}}\left[\frac{2\mathrm{Re}\left(e^{i\frac{1}{2}\pi\gamma}\right)}{(3\gamma-2)C_{CV1}^{1/2}}\left(\frac{k}{k_{CV1}}\right)^{\gamma-\frac{2}{3}}+1\right]^2 \tag{25}$$

Therefore, Eq. (21) gives

$$E_u(k) = C_{CV1}c_u^2k_{CV1}^{2\gamma-\frac{4}{3}}k^{-\frac{5}{3}}\left[\frac{2\mathrm{Re}\left(e^{i\frac{1}{2}\pi\gamma}\right)}{(3\gamma-2)C_{CV1}^{1/2}}\left(\frac{k}{k_{CV1}}\right)^{\gamma-\frac{2}{3}}+1\right]^2 \tag{26}$$

From Eqs. (25, 26), both the cascade of turbulent kinetic energy and scalar variance are dominated by the momentum diffusion features (e.g. $\gamma$ and $c_u$) in the CFV subrange. At the limit $k \ll k_{CV1}$, Eq. (25, 26) can be approximated to alternative forms as

$$E_s(k) \approx C_{CV1} N^2 k_{CV1}^{\frac{2}{3}-2\gamma} k^{2\gamma-\frac{11}{3}} \exp\left[\frac{4\mathrm{Re}\left(e^{i\frac{1}{2}\pi\gamma}\right)}{(3\gamma-2)C_{CV1}^{1/2}}\left(\frac{k}{k_{CV1}}\right)^{\gamma-\frac{2}{3}}\right] \tag{27a}$$

$$E_u(k) \approx C_{CV1} c_u^2 k_{CV1}^{2\gamma-\frac{4}{3}} k^{-\frac{5}{3}} \exp\left[\frac{4\mathrm{Re}\left(e^{i\frac{1}{2}\pi\gamma}\right)}{(3\gamma-2)C_{CV1}^{1/2}}\left(\frac{k}{k_{CV1}}\right)^{\gamma-\frac{2}{3}}\right] \tag{27b}$$

On the other hand, when $k \gg k_{CV1}$, Eq. (23) becomes

$$E_s(k) = k^{\frac{2\gamma-7}{3}}\left[C_{temp} - \frac{2N\mathrm{Re}\left(e^{i\frac{1}{2}\pi\gamma}\right)}{(1+\gamma)} k^{-\frac{1+\gamma}{3}}\right]^2 \tag{28}$$

Since Eqs. (6b) and (16b) are the same in both MFD and CFV subranges, it is reasonable to assume $E_s$ is also dominated by $\varepsilon_u$ and $\varepsilon_s$, as have been discussed in section 3. Therefore, the coefficient $C_{temp}^2$ associate to $E_s$ is dimensionally determined as $C_{CV2}\varepsilon_u^{(\gamma-2)/3}\varepsilon_s^{-(\gamma-2)}$, with $C_{CV2}$ being a dimensionless constant. Accordingly, Eq. (21) further gives

$$E_u(k) = C_{CV2}\varepsilon_u^{(\gamma-2)/3}\varepsilon_s^{-(\gamma-2)} M^2 c_u^{-2} k^{\frac{-4\gamma-1}{3}}\left[1 - \frac{2N\mathrm{Re}\left(e^{i\frac{1}{2}\pi\gamma}\right)}{C_{CV2}^{1/2}(1+\gamma)\varepsilon_u^{(\gamma-2)/6}\varepsilon_s^{-(\gamma-2)/2}} k^{-\frac{1+\gamma}{3}}\right]^2 \tag{29}$$

For ordinary momentum diffusion that $\gamma = 2$, they become

$$E_s(k) \approx \begin{cases} C_{CV1} N^2 k_{CV1}^{-10/3} k^{1/3} \exp\left[-C_{CV1}^{-1/2}\left(\dfrac{k}{k_{CV1}}\right)^{4/3}\right], & k_{MU} \ll k \ll k_{CV1} \\ C_{CV2} k^{-1}\left(1 + \dfrac{2N}{3C_{CV2}^{1/2}} k^{-1}\right)^2, & k_{CV1} \ll k \ll k_{CFVD} \end{cases} \tag{30a}$$

$$E_u(k) \approx \begin{cases} C_{CV1} c_u^2 k_{CV1}^{8/3} k^{-5/3} \exp\left[-C_{CV1}^{-1/2}\left(\dfrac{k}{k_{CV1}}\right)^{4/3}\right], & k_{MU} \ll k \ll k_{CV1} \\ C_{CV2} M^2 c_u^{-2} k^{-3}\left[1 + \dfrac{2N}{3C_{CV2}^{1/2}} k^{-1}\right]^2, & k_{CV1} \ll k \ll k_{CFVD} \end{cases} \tag{30b}$$

with $k_{CV1}|_{\gamma=2} = M^{1/3}N^{1/3}c_u^{-2/3}$. From Eqs. (30), it can be concluded that in the range $k_{MU} \ll k \ll k_{CV1}$, $\xi_u \approx -5/3$ and $\xi_s \approx 1/3$, indicating a slightly increasing of $E_s$ with $k$ (Fig. 5(a)). Beyond $k_{CV1}$, $\xi_u \approx -3$ and $\xi_s \approx -1$ (Fig. 5(b)). The CFV subrange terminates at $k_{CFVD}$, which is the smaller one between $k_{MZ} = (C_{CV2}Mc_u^{-2})^{1/\gamma}$ (the highest wavenumber that EBF cannot be negligible) and $k_{SZ} = (C_{CV2}Mc_u^{-1}c_s^{-1})^{1/\alpha}$ (the highest wavenumber that the scalar variance can be delivered by the convection induced by EBF). When $\gamma = \alpha = 2$, $k_{SZ} = k_{MZ}Sc_Z^{1/2}$. Regarding $Sc_Z \gg 1$, we finally get $k_{SZ} \gg k_{MZ} = k_{CFVD}$.

(2) On the other hand, if $\tau = \left(c_u^{2/\gamma}/\varepsilon_u\right)^{1/\left(3-\frac{2}{\gamma}\right)} = (c_u^2/\varepsilon_u^{\gamma})^{\frac{1}{(3\gamma-2)}}$ is the strain rate determined by momentum diffusion, according to Batchelor [22], furthermore we have

$$MNc_u^{-2}\frac{\mathrm{d}}{\mathrm{d}k}(k^{3-2\gamma}E_s)+k\frac{\mathrm{d}}{\mathrm{d}k}(E_s k)-2c_u^{\frac{4-3\gamma}{(3\gamma-2)}}\varepsilon_u^{\frac{\gamma}{(2-3\gamma)}}MN\mathrm{Re}\left(e^{i\frac{1}{2}\pi\gamma}\right)k^{2-\gamma}E_s=0 \quad (31)$$

Eq. (31) is also solved by Deepseek V4.1 and verified manually. The solution is

$$E_s(k)\sim k^{-1}(1+MNc_u^{-2}k^{1-2\gamma})^{-\frac{2(1-\gamma)}{1-2\gamma}}\exp\left[2c_u^{\frac{4-3\gamma}{(3\gamma-2)}}\varepsilon_u^{\frac{\gamma}{(2-3\gamma)}}MN\mathrm{Re}\left(e^{i\frac{1}{2}\pi\gamma}\right)\int\frac{k^{-\gamma}}{1+MNc_u^{-2}k^{1-2\gamma}}\,dk\right] \quad (32)$$

In the range $k_{MU}\ll k\ll k_{CV1}$, $MNc_u^{-2}k^{1-2\gamma}\gg 1$, regarding $(\varepsilon_u c_u^{-3})^{1/(3\gamma-2)}=k_K$ and $\gamma\geq 1$, we have

$$E_s(k)=C_{CV3}\,k_{CV1}^{2(1-\gamma)}k^{2\gamma-3}\exp\left[\frac{2\mathrm{Re}\left(e^{i\frac{1}{2}\pi\gamma}\right)}{\gamma}\left(\frac{k}{k_K}\right)^{\gamma}\right] \quad (33)$$

where $C_{CV3}$ is a dimensionless constant to be determined. Correspondingly, Eq. (21) gives

$$E_u(k)=C_{CV3}N^{-2}c_u^2k_{CV1}^{2\gamma}k^{-1}\exp\left[\frac{2\mathrm{Re}\left(e^{i\frac{1}{2}\pi\gamma}\right)}{\gamma}\left(\frac{k}{k_K}\right)^{\gamma}\right] \quad (34)$$

While in the range $k\gg k_{CV1}$, if $MNc_u^{-2}k^{1-2\gamma}\ll 1$ and $\gamma\neq 1/2$, thus

$$E_s(k)=\begin{cases}C_{CV4}k^{-1}\exp\left[\frac{2\mathrm{Re}\left(e^{i\frac{1}{2}\pi\gamma}\right)}{1-\gamma}\left(\frac{k_{CV1}}{k_K}\right)^{\gamma}\left(\frac{k}{k_{CV1}}\right)^{1-\gamma}\right], & \gamma\neq 1\\ C_{CV4}k^{-1}, & \gamma=1\end{cases} \quad (35)$$

Correspondingly, Eq. (21) gives

$$E_u=\begin{cases}C_{CV4}M^2c_u^{-2}k^{1-2\gamma}\exp\left[\frac{2\mathrm{Re}\left(e^{i\frac{1}{2}\pi\gamma}\right)}{1-\gamma}\left(\frac{k_{CV1}}{k_K}\right)^{\gamma}\left(\frac{k}{k_{CV1}}\right)^{1-\gamma}\right], & \gamma\neq 1\\ C_{CV4}M^2c_u^{-2}k^{-1}, & \gamma=1\end{cases} \quad (36)$$

For ordinary momentum diffusion that $\gamma=2$, they become

$$E_s(k)\approx\begin{cases}C_{CV3}\,k_{CV1}^{-2}k\exp\left[-\left(\frac{k}{k_K}\right)^2\right], & k_{MU}\ll k\ll k_{CV1}\\ C_{CV4}\,k^{-1}\exp\left[2\left(\frac{k_{CV1}}{k_K}\right)^2\left(\frac{k}{k_{CV1}}\right)^{-1}\right], & k_{CV1}\ll k\ll k_{CFVD}\end{cases} \quad (37a)$$

$$E_u(k)\approx\begin{cases}C_{CV3}N^{-2}c_u^2k_{CV1}^4k^{-1}\exp\left[-\left(\frac{k}{k_K}\right)^2\right], & k_{MU}\ll k\ll k_{CV1}\\ C_{CV4}N^{-2}c_u^2k_{CV1}^6k^{-3}\exp\left[2\left(\frac{k_{CV1}}{k_K}\right)^2\left(\frac{k}{k_{CV1}}\right)^{-1}\right], & k_{CV1}\ll k\ll k_{CFVD}\end{cases} \quad (37b)$$

In this case, we also predict two separated scaling regions below or above $k_{CV1}$, as shown in Fig. 5(b). When $k_{MU}\ll k\ll k_{CV1}$, $\xi_u\approx -1$ and $\xi_s\approx 1$, indicating a slightly increasing of $E_s$ with $k$ as well. When $k\gg k_{CV1}$, $E_u$ and $E_s$ may fall into either a scaling region with $\xi_u\approx -3$ and $\xi_s\approx -1$, or an exponential decaying region. In this case, $k_{MZ}=\left(C_{CV3}^{1/2}N^{-1}k_{CV1}^3\right)^{1/\gamma}$ and $k_{SZ}=\left(C_{CV3}^{1/2}N^{-1}c_uc_s^{-1}k_{CV1}^3\right)^{1/\alpha}$. When $\gamma=\alpha=2$, we still get $k_{SZ}=k_{MZ}Sc_Z^{1/2}$. Regarding $Sc_Z\gg 1$, $k_{SZ}\gg k_{MZ}=k_{CFVD}$.

In Fig. 5, we show the comparisons among Eqs. (30) in case (1) and Eqs. (37) in case (2) at $\gamma = \alpha = 2$, to demonstrate the influence of different time scales. In the range $k \ll k_{CV1}$, the two cases give different prediction on the scaling indices of $E_u$ and $E_s$. The case (1) predicts $E_u \sim k^{-5/3}$ and $E_s \sim k^{1/3}$, while the case (2) predicts $E_u \sim k^{-1}$ and $E_s \sim k$. The differences disappear in the high $k$ range ($k \gg k_{CV1}$), where both cases show $E_u \sim k^{-3}$ and $E_s \sim k^{-1}$. Unlike the case (1) where $E_u$ and $E_s$ in the low $k$ and high $k$ regions naturally connect, there is a gap in the case (2). This gap origins from two sources. One is from the different characteristic wavenumbers ($k_K$ and $k_{CV1}$) in Eqs. (33) and (35). The other is the difference between $C_{CV3}$ and $C_{CV4}$. For the latter, this gap can be smeared if $C_{CV4} = C_{CV3}\ \exp\left[2\mathrm{Re}\left(e^{i\frac{1}{2}\pi\gamma}\right)\left(\frac{1}{\gamma} - \frac{1}{1-\gamma}\right)\left(\frac{k_{CV1}}{k_K}\right)^{\gamma}\right]$. While for the former, detailed intrinsic relationship among $M$, $N$, $c_u$, $c_s$, $\varepsilon_u$ and $\varepsilon_s$ must be established in advance regarding $\gamma$ and $\alpha$. Unfortunately, this was not obtained in the current investigation.

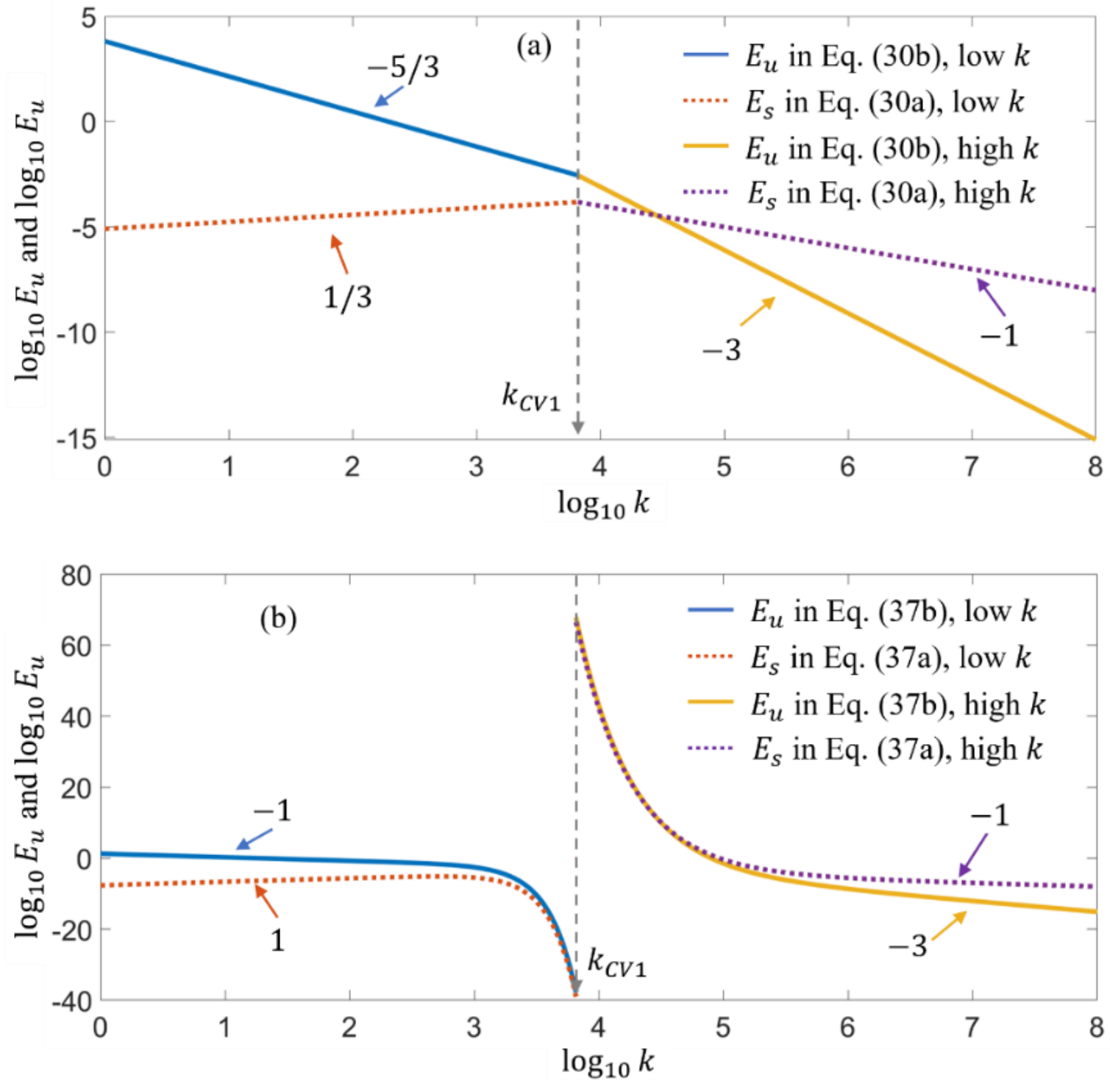

Fig. 5 An example of EK turbulence in liquid phase, where the VF subrange is connected to the CFV subrange. Here, $\varepsilon = \varepsilon_0 \varepsilon_r$, $\varepsilon_0 = 8.854 \times 10^{-12}$ F/m, $\varepsilon_r = 80$, $E_y = 10^5$ V/m, $\langle\sigma\rangle_0 = 0.25$ S/m, $M = \varepsilon E_y^2/\rho\langle\sigma\rangle_0 = 0.0283$ m$^2$/s$^2$, $N = |\nabla\langle\sigma\rangle| = 10$ S/m$^2$, $\rho = 10^3$ kg/m$^3$, $c_u = 10^{-6}$ m$^2$/s$^3$, $c_s = 10^{-9}$ 1/s, $\gamma = \alpha = 2$. (a) $E_u$ and $E_s$ in the CFV subrange of case (1). $C_{CV1} = 4.31 \times 10^5$, $C_{CV2} = 1$, $k_{CV1} = 6.57 \times 10^3$ 1/m, $k_{CFVD} = 1.68 \times 10^5$ 1/m. (b) $E_u$ and $E_s$ in the CFV subrange of case (2). $C_{CV3} = C_{CV4} = 1$, $k_{CV1} = 6.57 \times 10^3$ 1/m, $k_K = 7.30 \times 10^2$ 1/m, $k_{CFVD} = 1.68 \times 10^5$ 1/m.

### *6.2 In the case $Sc_Z \ll 1$*

Similar as in Part 1 [12], there exist two possible cases when $Sc_Z \ll 1$. Since there are Quad-cascade sub-branches with different scaling properties in the spectra of turbulent kinetic energy and scalar, the situations become ultracomplex when they interconnect with the dissipation subranges. Here, as examples, we only consider the case when the inertial or constant-$\Pi_u$ sub-branches of the MFD subrange are interconnected with the dissipation subrange.

(1) When $\gamma > \alpha$, for instance, $\gamma = 2$ and $\alpha = 3/2$, we have $k_{SD} = Sc_Z^{6/5}(k_0/k_{MD})^{-3/5}k_{MD}$. Even though $Sc_Z \ll 1$, we can still get $k_{SD} \gg k_{MD}$ if $k_0/k_{MD}$ is sufficiently small. At wavenumber $k = k_{MD}$, the time required for momentum diffusion (i.e. $(c_u k_{MD}^{\gamma})^{-1}$ ) is much shorter than that of scalar diffusion (i.e. $(c_s k_{MD}^{\alpha})^{-1} = (k_0/k_{MD})^{\alpha-\gamma} Sc_Z (c_u k_{MD}^{\gamma})^{-1}$). Thus, the scalar structure can still be transported into a larger wavenumber (or smaller scale) relative to velocity structure, even if $Sc_Z \ll 1$. For this case, the CFV subrange is still predictable.

(2) When $\gamma \leq \alpha$, for instance, $\gamma = 3/2$ and $\alpha = 2$, $k_{SD} = Sc_Z^{3/4}(k_0/k_{MD})^{3/8}k_{MD}$ if the inertial or constant-$\Pi_u$ sub-branches of the MFD subrange are interconnected with the dissipation subrange. Since both $Sc_Z$ and $k_0/k_{MD}$ are much smaller than unity, we have $k_{SD} = k_{MU} \ll k_{MD}$. Thus, there could exist DFC subrange (Fig. 4) in the range $k_{MU} \ll k \ll k_{DFCD}$, with $k_{DFCD}$ being the upper bound of DFC subrange. In this subrange, the scalar structures in the scalar dissipation subrange determine the electric body force, which in turn drives the flow, induces scalar convection and finally feedback to scalar dissipation subrange.

In this subrange,

$$\frac{\mathrm{d}}{\mathrm{d}k}\Pi_u(k) = F_s(k) \tag{38a}$$

$$\frac{\mathrm{d}}{\mathrm{d}k}\Pi_s(k) = -F_A(k) + D_s(k) \tag{38b}$$

Regarding $F_s(k) - \frac{M}{N}F_A(k)k^{\beta} = 0$,

$$\frac{\mathrm{d}}{\mathrm{d}k}\Pi_u(k) + \frac{M}{N}\left[\frac{\mathrm{d}}{\mathrm{d}k}\Pi_s(k) - D_s(k)\right]k^{\beta} = 0 \tag{39}$$

which can be further rewritten as

$$\frac{\mathrm{d}}{\mathrm{d}k}\left(\frac{E_u k}{\tau}\right) + \frac{M}{N}k^{\beta}\frac{\mathrm{d}}{\mathrm{d}k}\left(\frac{E_s k}{\tau}\right) - 2\frac{M}{N}c_s \mathrm{Re}\left(e^{i\frac{1}{2}\pi\alpha}\right)k^{\alpha+\beta}E_s = 0 \tag{40}$$

In the following part, we only consider the case $\beta = 1$. To establish the relationship between $E_u$ and $E_s$, we employ a similar phenomenological treatment as the inertial-diffusive subrange [12,23], i.e.

$$c_s^2 k^{2\alpha}E_s \sim E_u G_s \tag{41}$$

where $G_s = \langle|\nabla\widehat{s'}|^2\rangle$ is the mean module of scalar gradient. Here, we use small-scale scalar time $\tau = (c_s^2/\varepsilon_u^{\alpha})^{\frac{1}{3\alpha-2}}$, as the characteristic time scale to evaluate scalar diffusion, then Eq. (35) becomes

$$c_s^2 G_s^{-1}\frac{\mathrm{d}}{\mathrm{d}k}(k^{2\alpha+1}E_s) + \frac{M}{N}k\frac{\mathrm{d}}{\mathrm{d}k}(E_s k) - 2\frac{M}{N}\mathrm{Re}\left(e^{i\frac{1}{2}\pi\alpha}\right)c_s^{\frac{3\alpha}{3\alpha-2}}\varepsilon_u^{-\frac{\alpha}{3\alpha-2}}k^{\alpha+1}E_s = 0 \tag{42}$$

After solving this equation by Deepseek V4.1, we have

$$E_s(k) \sim k^{-2}NM^{-1}(NM^{-1}c_s^2 G_s^{-1}k^{2\alpha-1}+1)^{-1}\exp\left[\int \frac{1+2\mathrm{Re}\left(e^{i\frac{1}{2}\pi\alpha}\right)c_s^{\frac{3\alpha}{3\alpha-2}}\varepsilon_u^{-\frac{\alpha}{3\alpha-2}}k^{\alpha}}{NM^{-1}c_s^2 G_s^{-1}k^{2\alpha-1}+1}k^{-1}dk\right] \tag{43}$$

In the range $\alpha \geq 1$, when $k \gg k_{CV2} = (NM^{-1}c_s^2 G_s^{-1})^{-1/(2\alpha-1)}$ which is a characteristic scalar wavenumber regarding forcing, $NM^{-1}c_s^2 G_s^{-1} k^{2\alpha-1} \gg 1$, then

$$E_s(k) \sim \begin{cases} c_s^{-2} G_s k^{-1-2\alpha} \exp\left[\frac{1}{1-2\alpha}\left(\frac{k}{k_{CV2}}\right)^{1-2\alpha} + \frac{2\mathrm{Re}\left(e^{i\frac{1}{2}\pi\alpha}\right)}{1-\alpha}\left(\frac{k_{CV2}}{k_S}\right)^{\alpha}\left(\frac{k}{k_{CV2}}\right)^{1-\alpha}\right], \alpha \neq 1 \\ c_s^{-2} G_s k^{-1-2\alpha} \exp\left[-\left(\frac{k}{k_{CV2}}\right)^{-1}\right], \quad \alpha = 1 \end{cases} \tag{44}$$

According to Eq. (41),

$$E_u(k) \sim \begin{cases} k^{-1} \exp\left[\frac{1}{1-2\alpha}\left(\frac{k}{k_{CV2}}\right)^{1-2\alpha} + \frac{2\mathrm{Re}\left(e^{i\frac{1}{2}\pi\alpha}\right)}{1-\alpha}\left(\frac{k_{CV2}}{k_S}\right)^{\alpha}\left(\frac{k}{k_{CV2}}\right)^{1-\alpha}\right], \alpha \neq 1 \\ k^{-1} \exp\left[-\left(\frac{k}{k_{CV2}}\right)^{-1}\right], \quad \alpha = 1 \end{cases} \tag{45}$$

Since Eq. (38a) is the same in both MFD and DFC subranges, it is reasonable to assume $E_u$ is dominated by $\varepsilon_u$ and $\varepsilon_s$, as have been discussed in section 3. Therefore, the coefficient associate to $E_u$ and $E_s$ is $C_{CV5}\varepsilon_u\varepsilon_s^{-1}$, with $C_{CV5}$ being a dimensionless constant.

When $k \ll k_{CV2}$, $NM^{-1}c_s^2 G_s^{-1} k^{2\alpha-1} \ll 1$, we have

$$E_s(k) \sim NM^{-1}\, k^{-1} \exp\left\{\frac{2\mathrm{Re}\left(e^{i\frac{1}{2}\pi\alpha}\right)}{\alpha}\left(\frac{k}{k_S}\right)^{\alpha}\right\} \tag{46}$$

According to Eq. (41) as well,

$$E_u(k) \sim \left(\frac{k}{k_{CV2}}\right)^{2\alpha-1} \exp\left\{\frac{2\mathrm{Re}\left(e^{i\frac{1}{2}\pi\alpha}\right)}{\alpha}\left(\frac{k}{k_S}\right)^{\alpha}\right\} \tag{47}$$

The coefficient associate to $E_u$ and $E_s$ in Eqs. (46) and (47) can be $C_{CV6}\varepsilon_u^{3/2}\varepsilon_s^{-5/2}$, with $C_{CV6}$ being another dimensionless constant to be determined.

For ordinary scalar diffusion that $\alpha = 2$, they become

$$E_s(k) = \begin{cases} C_{CV6}\varepsilon_u^{3/2}\varepsilon_s^{-5/2} NM^{-1}\, k^{-1} \exp\left\{-\left(\frac{k}{k_S}\right)^2\right\}, & k_{MU} \ll k \ll k_{CV2} \\ C_{CV5}\varepsilon_u\varepsilon_s^{-1} c_s^{-2} G_s k^{-5} \exp\left[-\frac{1}{3}\left(\frac{k}{k_{CV2}}\right)^{-3} + 2\left(\frac{k_{CV2}}{k_S}\right)^2\left(\frac{k}{k_{CV2}}\right)^{-1}\right], & k_{CV2} \ll k \ll k_{DFCD} \end{cases} \tag{48a}$$

$$E_u(k) = \begin{cases} C_{CV6}\varepsilon_u^{3/2}\varepsilon_s^{-5/2} \left(\frac{k}{k_{CV2}}\right)^3 \exp\left\{-\left(\frac{k}{k_S}\right)^2\right\}, & k_{MU} \ll k \ll k_{CV2} \\ C_{CV5}\varepsilon_u\varepsilon_s^{-1} k^{-1} \exp\left[-\frac{1}{3}\left(\frac{k}{k_{CV2}}\right)^{-3} + 2\left(\frac{k_{CV2}}{k_S}\right)^2\left(\frac{k}{k_{CV2}}\right)^{-1}\right], & k_{CV2} \ll k \ll k_{DFCD} \end{cases} \tag{48b}$$

In this case, we also predict two separated scaling regions below or above $k_{CV2}$, as shown in Fig. 6. When $k_{MU} \ll k \ll k_{CV2}$, $\xi_u \approx 3$ and $\xi_s \approx -1$, indicating a slightly increasing of $E_u$ with $k$. When $k_{CV2} \ll k \ll k_{DFCD}$, $E_u$ and $E_s$ may fall into either a scaling region with $\xi_u \approx -1$ and $\xi_s \approx -5$, or an exponential decaying region. There is also a gap in Fig. 6 around $k = k_{CV2}$. This gap origins from either the different characteristic wavenumbers ($k_S$ and $k_{CV2}$) in Eqs. (48), or the difference between $C_{CV5}$ and $C_{CV6}$. For the latter, this gap can be smeared if $C_{CV5} = C_{CV6}\varepsilon_u^{1/2}\varepsilon_s^{-3/2}NM^{-1}c_s^2G_s^{-1}k_{CV2}^{2\alpha}\exp\left\{\left(\frac{1}{\alpha}+\frac{1}{\alpha-1}\right)2\mathrm{Re}\left(e^{i\frac{1}{2}\pi\alpha}\right)\left(\frac{k_{CV2}}{k_S}\right)^{\alpha}-\frac{1}{1-2\alpha}\right\}$.

In this case, $k_{DFCD}$ equals to the smaller one between $k_{MZ}$ and $k_{SZ}$. They do not have an explicit solution, but should be solved numerically from $k_{MZ} = [k_{MZ}E_u(k_{MZ})]^{\frac{1}{2(\gamma-1)}}c_u^{-\frac{1}{\gamma-1}}$ and $k_{SZ} = [k_{SZ}E_u(k_{SZ})]^{\frac{1}{2(\alpha-1)}}c_s^{-\frac{1}{\alpha-1}}$ respectively, where $E_u$ should take the second solution in Eq. (48b). Theoretically, it should be predicted $k_{SZ} = k_{DFCD} \ll k_{MZ}$.

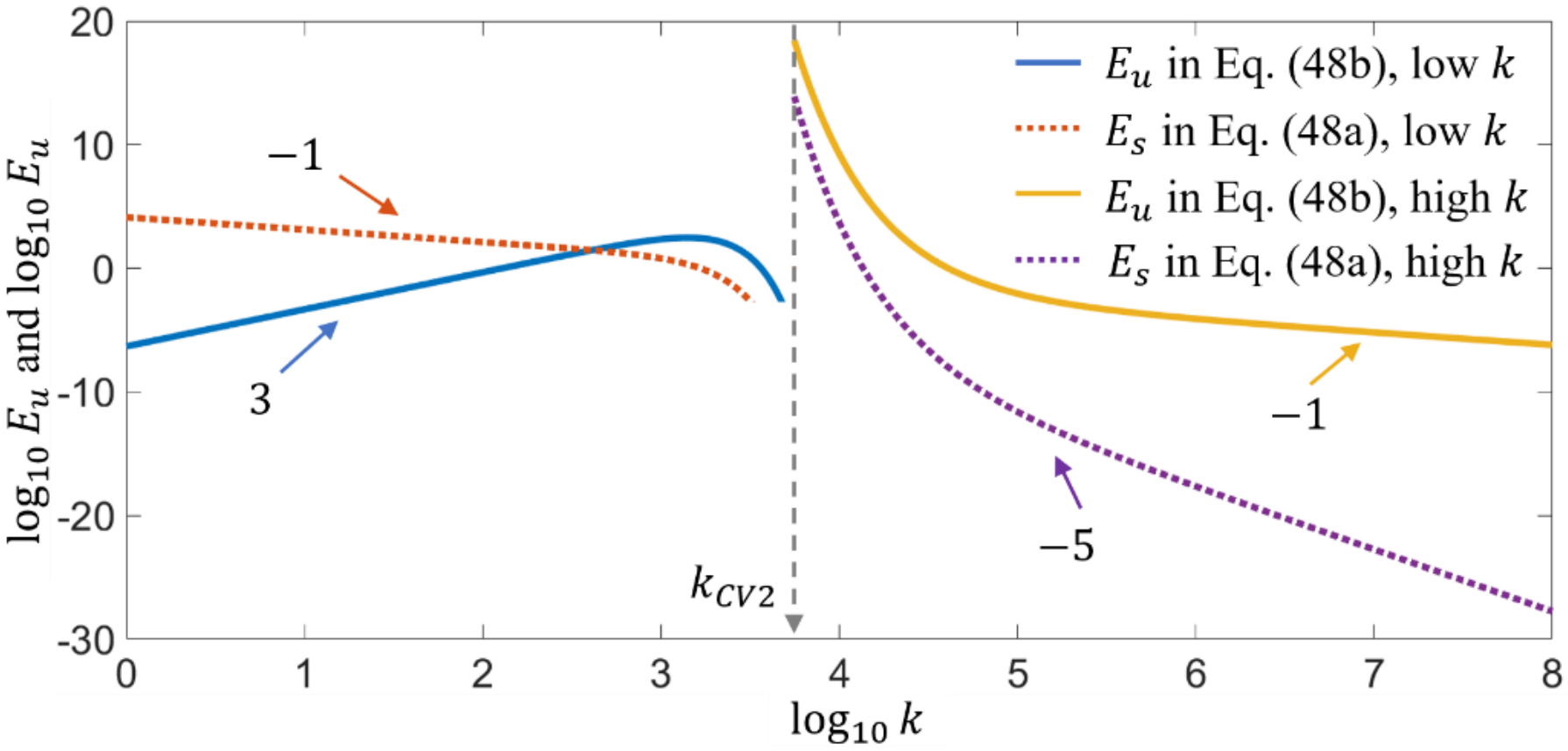

Fig. 6 An example in plasma turbulence driven by external electric field, regarding the VF subrange is connected to the DFC subrange. Here, $\varepsilon = \varepsilon_0\varepsilon_r$, $\varepsilon_0 = 8.854 \times 10^{-12}$ F/m, $\varepsilon_r = 80$, $\langle\sigma\rangle_0 = 0.1$ S/m, $M = \varepsilon E_y^2/\rho\langle\sigma\rangle_0 = 62.7$ m$^2$/s$^2$, $N = |\nabla\langle\sigma\rangle| = 10$ S/m$^2$, $E_y = 10^5$ V/m, $\rho = 1.13$ kg/m$^3$, $c_u = 1.53 \times 10^{-6}$ m$^2$/s$^3$, $c_s = 6 \times 10^{-5}$ 1/s (inferred from Ref. [24]), $\gamma = \alpha = 2$, $G_s = 10^2$ 1/m$^2$. $C_{CV5} = C_{CV6} = 1$, $k_{CV2} = 5.58 \times 10^3$ 1/m, $k_K = 1.80 \times 10^4$ 1/m, $k_S = 1.15 \times 10^3$ 1/m, $k_{DFCD} = k_{SZ} = 1.68 \times 10^5$ 1/m, $k_{MZ} = 5.30 \times 10^6$ 1/m.

## 7. Discussion and conclusions

To this end, we have established a comprehensive picture for velocity and scalar cascades in momentum–scalar coupled turbulence under short-range forcing ($\beta > 2/3$), accounting for anomalous diffusion of both momentum and scalar. This Part 2 extends the generalized framework developed in Part 1 to the regime where the MFD subrange lies on the high-wavenumber side of the inertial subrange, and the Quad-cascade process model fundamentally alters the cascade pathways.

Three main contributions can be summarized from our analysis. First, we derived the general relations for $k_{MD}$ and $k_{SD}$ as functions of the scaling exponents $\xi_u$ and $\xi_s$, which depend on the specific Quad-cascade sub-branch. These relations show that $k_{MD}$ and $k_{SD}$ are not independent, but are entangled. Second, by specializing to EK turbulence ($\beta = 1$), we demonstrated how the three anomalous diffusion regimes of electrolytes ($\alpha = 1, 4/3, 3/2$) map onto the Quad-cascade sub-branches, revealing that the cascade topology is highly sensitive to which regime the dissipation range falls into. Third, we identified two new spectral subranges — the CFV subrange and the DFC subrange — that exist exclusively for short-range forcing. The former exists in $k_{MU} \ll k \ll k_{CFVD}$ at either $Sc_Z \gg 1$ or $Sc_Z \ll 1$ with $\gamma > \alpha$, where the scalar field drives viscous flow through electric body force. While the latter exists in $k_{MU} \ll k \ll k_{DFCD}$ at $Sc_Z \ll 1$ and $\gamma \leq \alpha$, where the scalar structure dominated by diffusion determines electric body force, which in turn cause momentum convection. These subranges have no counterpart in Part 1 or in classical theories, representing genuinely new physics arising from the interplay between short-range forcing and anomalous diffusions. We have summarized the subranges in Table 5 for comparison.

Table 5. Comparisons among the subranges in both Part 1 and 2.

| Subrange | Physical meaning | Properties |
|---|---|---|
| Viscous-convective | Viscous effect dominates momentum transport, which in turn controls scalar transport by convection | $E_s \sim k^{-1} \exp\left[2\mathrm{Re}\left(e^{i\frac{1}{2}\pi\alpha}\right)\alpha^{-1}\left(\frac{k}{k_B}\right)^{\alpha}\right]$ |
| Inertial-diffusive | Inertial effect dominates momentum transport, which in turn controls scalar transport by diffusion | $E_s \sim \left(\frac{k}{k_{S3}}\right)^{-\frac{5}{3}-2\alpha} + \left(\frac{k}{k_{S4}}\right)^{\gamma-2\alpha-\frac{7}{3}} + \left(\frac{k}{k_{S5}}\right)^{2\gamma-2\alpha-3}$ |
| Convective-forcing-viscous | Scalar convection dominates the evolution of scalar structure, generating electric body force simultaneously, which in turn induces viscous flow | Eqs. (27-29) or Eqs. (33-36) |
| Diffusive-forcing-convective | Scalar structures evolve according to scalar diffusion, generating electric body force simultaneously, which in turn induces momentum convection | Eqs. (44-47) |

In both the CFV and DFC subranges, the Quad-cascade branches predicted in the MFD subrange are not observed. Instead, we predict unique $E_u$ and $E_s$ under the given time scales. This indicates a bifurcation of the dynamical system, from multiple states to a single state. The dissipation terms $D_u$ and $D_s$ in the spectral budget equations of $\Pi_u - \Pi_s$ pair explicitly break the scale invariance that holds in the MFD subrange. This explicit symmetry breaking changes the degeneracy among the Quad-cascade branches and selects a unique spectral solution in the CFV and DFC subranges, implying the important influence of the diffusions of momentum and scalar. In fact, the diffusions of momentum (e.g. viscosity) affecting the symmetry of physical system has been found in fluid dynamics [25,26] and plasma physics [27]. In this model, we demonstrate not only momentum diffusion, but also scalar diffusion can cause explicit symmetry breaking of scaling invariance.

In the meanwhile, several limitations should be acknowledged as well. As discussed in Part 1, some predicted characteristic wavenumbers, particularly for $\alpha = 1$ in the Constant-$\Pi_\mathrm{s}$ and VF subranges, may exceed the physical bounds of the continuum hypothesis (Fig. 3). This suggests either that such extreme anomalous exponents are physically unrealizable in real electrolytes, or that additional physical mechanisms (e.g., Debye screening, finite ion size, or quantum effects) become relevant before such scales are reached.

To the best of our knowledge, no direct numerical simulations (DNS) or experimental data are currently available for EK turbulence with anomalous diffusion at the level of detail required to test the present predictions. However, the scaling laws and phase diagrams provided here offer specific, testable predictions for future numerical and experimental studies. In particular, the predicted transitions between the convective-forcing-viscous and diffusive-forcing-convective subranges as functions of $Sc_Z$—controlled externally by electric field strength and ionic diffusivity—provide clear signatures that can be sought in laboratory EK flows.

In summary, Parts 1 and 2 together establish a comprehensive theoretical framework that unifies anomalous diffusion and multiscale forcing in momentum–scalar coupled turbulence, revealing previously unrecognized effects of fractional transport on turbulent cascades across both long-range and short-range forcing regimes. The derived scaling laws and characteristic length scales offer testable predictions for a wide range of natural and engineering disciplines where ordinary diffusion fails to capture the small-scale dynamics.

**Acknowledgement** The investigation is supported by the Open Project of the Shaanxi Provincial Key Laboratory of Optoelectronic Technology (SXPEL2026O-02).